\documentclass[onecolumn,tighten]{aastex62}
\usepackage{amsmath} \usepackage{amssymb}
\usepackage{CJKutf8}
\usepackage{lineno}
\hypersetup{linkcolor=red,citecolor=cyan,filecolor=cyan,urlcolor=magenta}

\usepackage{natbib}
\newcommand{{\HII}}{H\,{\sc ii} }
\newcommand{{\HI}}{H\,{\sc i} }

\def\degree{${}^{\circ}$}
\usepackage{longtable}
\usepackage{threeparttablex}
\usepackage{makecell}

\received{June 28, 2022}
\revised{October 31, 2022}
\accepted{November 12, 2022}
\submitjournal{ApJS}

\shorttitle{WISE Green Objects}
\shortauthors{Zhang et al.}

\usepackage{longtable,booktabs}
\begin{document}

\title{WISE Green Objects (WGOs): the massive star candidates in
		the whole Galactic Plane ($\mid b \mid <2$\degree) }

\correspondingauthor{Chang Zhang, Guo-Yin Zhang, Jin-Zeng Li}

\author{Chang Zhang}
\affil{National Astronomical Observatories,
	Chinese Academy of Sciences,
	A20 Datun Road, Chaoyang District,
	Beijing 100101, PR China; zhangc@nao.cas.cn, zgyin@nao.cas.cn, ljz@nao.cas.cn}
\affil{University of Chinese Academy of Sciences, Beijing 100049, PR China}

\author{Guo-Yin Zhang}
\affil{National Astronomical Observatories,
	Chinese Academy of Sciences,
	A20 Datun Road, Chaoyang District,
	Beijing 100101, PR China; zhangc@nao.cas.cn, zgyin@nao.cas.cn, ljz@nao.cas.cn}
	
\author{Jin-Zeng Li}
\affil{National Astronomical Observatories,
	Chinese Academy of Sciences,
	A20 Datun Road, Chaoyang District,
	Beijing 100101, PR China; zhangc@nao.cas.cn, zgyin@nao.cas.cn, ljz@nao.cas.cn}
	
\author{Jing-Hua Yuan}
\affil{National Astronomical Observatories,
	Chinese Academy of Sciences,
	A20 Datun Road, Chaoyang District,
	Beijing 100101, PR China; zhangc@nao.cas.cn, zgyin@nao.cas.cn, ljz@nao.cas.cn}

\begin{abstract}
Massive young stellar objects (MYSOs) play a crucial role in star formation. Given that MYSOs were previously identified based on the extended structure and the observational data for them is limited, screening the Wide-field Infrared Survey Explorer (WISE) objects showing green features (for the common coding of the 4.6 $\mu$m band as green channel in three-color composite WISE images) will yield more MYSO candidates. Using WISE images in the whole Galactic Plane ($ 0^\circ<l<360^\circ $ and $\mid b \mid <2$\degree), we identified sources with strong emissions at 4.6 $\mu$m band, then according to morphological features divided them into three groups. We present a catalog of 2135 WISE Green Objects (WGOs). 264 WGOs have an extended structure. 1366 WGOs show compact green feature but without extended structure. 505 WGOs have neither extended structure nor green feature, but the intensity at 4.6 $\mu$m is numerically at least 4.5 times that of 3.4 $\mu$m. According to the analysis of the coordinates of WGOs, we find WGOs are mainly distributed in $\mid l \mid<$ 60\degree, coincident with the position of the giant molecular clouds in $\mid l \mid>$ 60\degree. Matching results with various masers show that those three groups of WGOs are at different evolutionary stages. After cross-matching WGOs with published YSO survey catalogs, we infer that $\sim$50\% of WGOs are samples of newly discovered YSOs. In addition, 1260 WGOs are associated with Hi-GAL sources, according to physical parameters estimated by spectral energy distribution fitting, of which 231 are classified as robust MYSOs and 172 as candidate MYSOs.
\end{abstract}

\keywords{ISM: WISE Green Objects ---
	ISM: Outflow --
	stars: formation --- stars: massive---stars: protostars}

\section{Introduction} \label{sec:intro}
As the main contributor to emission and chemical enrichment of the universe, the formation of stars has invariably been an important research topic in astronomy. The scenario of low-mass star formation has been well established \citep{Shu1987}. Statistics suggest that massive star formation is unlikely to be found nearby, with most being $>$ 2 kpc away. Influenced by the factors such as star-forming regions are rare, deeply embedded, the timescales are extremely short, and the natal environments are inevitably destroyed by violent feedback, the formation process of the massive star ($\gtrsim 8\ M_{\odot}$) is still a mystery \citep{Zinnecker2007}. The result of these observational obstacles is that there are few accurate or well-selected samples of objects in the early stages of evolution. A large, unbiased sample of massive young stellar objects (MYSOs), especially those in infancy, could help understand the processes involved in the formation and earliest stages of massive star formation \citep{Urquhart2014,Urquhart2022}.

The “monolithic collapse” model \citep{2002Natur.416...59M,2003ApJ...585..850M} and the “competitive accretion” model \citep{1997MNRAS.285..201B,2004sptz.prop.3394B} are two popular models of massive star formation. The “monolithic collapse” model is an extended version of low-mass star formation, in which gas is accreted onto the protostar through an accretion disk at a significantly higher rate of accretion compared to its low-mass star formation. The final stellar mass comes from the initial core mass in this model. The “competitive accretion” model is that massive stars always form in clusters and rely on the competitive accumulation of cluster members from a common envelope. The final stellar mass depends on the result of the competition. Regardless of the mass, there is evidence of the same nature of low- and high-mass star formation. For example, the latest findings of accretion bursts in MYSO \citep[e.g.,][]{Caratti2017,2017arXiv171201451S,2021A&A...646A.161S} reveal that both low- and high-mass protostars form through disk accretion, accompanied by episodic accretion bursts, possibly caused by disk fragmentation. The typical the accretion rate during the low-mass star formation is $\sim 5\times 10^{-6} \ M_{\odot }\rm\ yr^{-1}$ \citep{Hosokawa2010}. In estimating this value, a typical dust temperature of 10 K in the cold core is adopted \citep{Zhang2022}. This value during massive star formation is expected to be $\gtrsim 10^{-4} \ M_{\odot }\rm\ yr^{-1}$ \citep{Hosokawa2010}. Such high accretion rates support the ability of lower-mass progenitors have the ability to accrete enough material from their gas-rich circumstellar disks to grow into massive stars (e.g., \citealt{Motte2018,2021ApJ...922...90C}).

The MYSO samples were determined mainly by observational data provided by {\it IRAS} \citep{Molinari1996} and the $\it Midcourse~Space~Experiment$ ($\it MSX$) \citep{Hoare2005}, while the {\it Spitzer} surveys of the Galactic Plane using the  InfraRed Array Camera (IRAC, 3.6, 4.5, 5.8 and 8.0 $\mu$m; \citealt{2004ApJS..154...10F}) that came later replaced the former with sub-2\arcsec \ resolution and higher sensitivity. The  {\it Spitzer} Galactic Legacy Infrared Mid-Plane Survey Extraordinaire (GLIMPSE, \citealt{2001AAS...198.2504C}) I / II has revealed several ``Extended Green Objects" (EGOs) which display extended emission in the 4.5 $\mu$m band coded as the green channel in the trichromatic image. The extended green emission is thought to be caused by MYSO outflows \citep{Cyganowski2008, Chen2013}. When the material is continually accreted from the disk to the protostar, it will release excess angular momentum and produce collimated jets or outflows (e.g., \citealt{Shu1987, Hosokawa2010, Motte2018, 2021ApJ...922...90C}). The {\it Spitzer} 4.5 $\mu$m band covers excitation radiation of H$_{2}$ ($v$ = 0 – 0, S (9, 10, 11)) and CO ($v$ = 1 – 0) that forms in the regions of interaction between outflows and interstellar medium \citep{2005MNRAS.357.1370S,2006AJ....131.1479R,Davis2007}. The sources with extended 4.5 $\mu$m emission have a high percentage of shock-triggered masers (e.g., H$_2$O and CH$_{3}$OH Class I, \citealt{Cyganowski2009, Cyganowski2013, Chen2011, Towner2017}). As the signposts of MYSOs, class II methanol masers have a high detection rate in MYSOs \citep{Jones2020,2021A&A...646A.161S}. 
The above evidences show that the EGOs are excellent MYSO candidates with active outflows.
%are considered to be excited by shock waves generated by outflow interact with interstellar medium during star formation

The {\it Spitzer} observation range is limited to the inner Galactic Plane ($\mid l \mid<$ 65\degree, \citealt{Churchwell2009}). The observation area of Wide-field Infrared Survey Explorer (WISE) covers the entire Galactic Plane \citep{Wright2010}. The exposure time of {\it Spitzer} GLIMPSE is 4 seconds per frame, and 5$\sigma$ sensitivity is $\sim $ 0.2 mJy at 4.5 $\mu$m band \citep{2003PASP..115..953B}. The exposure time of WISE is 11 seconds per frame, and 5$\sigma$ sensitivity is $\sim $ 0.11 mJy at 4.6 $\mu$m band \citep{Wright2010}. Although WISE has an angular resolution about 3 times smaller than that of {\it Spitzer}, but thanks to the long exposure time of WISE, the sensitivity of WISE is 2 times higher than that of {\it Spitzer}. In addition, the WISE data we adopted is $\sim$1440 deg$^{2}$, which is five times of {\it Spitzer} GLIMPSE ($\sim$274 deg$^{2}$). Considering sample size = (solid angle)*N($>$S) where N($>$S) is the surface number density of sources brighter than a flux limit S, and N($>$S) $\sim $ 1/S, WISE is expected to find ten times as many sources as {\it Spitzer}. 

\cite{Cyganowski2008} and \cite{Chen2013} identified MYSOs by checking whether the infrared sources have extended structures. In this work, WISE W1, W2, and W3 bands are encoded as blue, green, and red, respectivley, in the three-color composite images. We call sources that appear green in color or sources whose W2 is significantly larger than W1 in value as WGOs. Dense cores are localized density enhancements of the cloud material that have been recognized for sites of low- and high mass star formation for more than 30 years \citep{2007ARA&A..45..339B}. WGOs, which do not show 4.6 $\mu$m extended emission, are still deeply embedded in the cores and may be at an early evolutionary stage of MYSOs. As long as there is enough gas on the envelope or disk, they have the potential to grow into MYSOs with extended structures \citep{1996ApJ...462..874J,2002ApJ...569..846Y}. The ultra-low temperature cooled cameras carried by {\it Herschel} can detect far-infrared dust radiation from the dense cores \citep{2010A&A...518L...1P}. Based on the size and mass of the core, it can be inferred whether the core can give birth to massive stars (e.g., \citealt{2008Natur.451.1082K,2010ApJ...723L...7K,Zhang2018}). The emission of YSOs in different bands will change with the evolution process and this change is reflected in the shape of the spectrum energy distribution (SED) fitting (e.g., \citealt{Shu1987,1987IAUS..115....1L,1993ApJ...406..122A,1994ApJ...434..614G}). By fitting the SED of YSOs with the corresponding theoretical model \citep{Robitaille2017}, we can obtain various property parameters of YSOs, such as mass, temperature, and radius. By combining these parameters, the accretion rate of YSOs can be estimated, which can predict whether YSOs can form massive stars (e.g., \citealt{2020ApJ...897..136M,2019MNRAS.490..832J,2021A&A...646A.161S}).
%WGOs, still deeply embedded in the cores, may be at an early evolutionary stage of MYSOs with extended structures.

The outline of the paper is as follows. In Sect. \ref{sec:2}, we describe archived data used for this article. Data analysis and results are presented in Sect. \ref{Dataanalysis}. Sect. \ref{DISCUSSION} discusses star formation scenario of WGOs and reliability of WGOs as MYSOs. We summarize our conclusions in Sect. \ref{CONCLUSIONS}.

\section{Archive data}\label{sec:2}

\subsection{WISE data}\label{sec:2.1}
WISE, equipped with a 40 cm diameter infrared telescope, performed an all-sky astronomical survey in Earth orbit over ten months in four infrared bands, W1, W2, W3, and W4, centered at 3.4, 4.6, 12, and 22 $\mu$m wavelength \citep{Wright2010}. The angular resolutions are 6.1\arcsec, 6.4\arcsec, 6.5\arcsec, and 12.0\arcsec\ respectively for wavelengths of 3.4, 4.6, 12, and 22 $\mu$m, and in unconfused regions on the ecliptic point source sensitivities at 5$\sigma$ are better than 0.08, 0.11, 1, and 6 mJy. The data were released on 14 March 2012 and can be retrieved from the Infrared Science Archive\footnote{\url{https://irsa.ipac.caltech.edu/Missions/wise.html}}. We selected all the data in the range of the Galactic Plane ($\mid b \mid <2$\degree) covering $\sim $ 1440 deg$^{2}$, including a total of $\sim $600 small images with the size of 1.56\degree $\times$ 1.56\degree.

\subsection{Hi-GAL data}\label{sec:2.2}
{\it Herschel} infrared Galactic Plane Survey (Hi-GAL) was performed in five infrared continuum bands between 70 and 500 $\mu $m to map the dust distribution in $\mid b \mid <$ 1.5\degree~ \citep{Molinari2010}, that provide a census of dense and cold condensations that some sources may have harbored YSOs. \cite{Mege2021} resolved distances for $\sim $80\% compact sources by substituting radial velocity into the rotation of the Galaxy and assisted with \HI self-absorption method or distance–extinction data to solve distance ambiguity. Based on the fitting of a modified black-body (grey-body) function to {\it Herschel} $\lambda \geqslant 160~\mu m$ portion and the distances given by \cite{Mege2021}, \cite{2021MNRAS.504.2742E} derived source physical properties, including the mass $M$, the temperature $T$, and bolometric luminosity $L$ and so on. The source size is measured from 250 $\mu $m wavelength image.

\section{Data analysis and results} \label{Dataanalysis}
\subsection{Identification and Classification of WGOs}\label{sec:3.1}

\begin{deluxetable*}{p{7cm}|p{7cm}}
	\tablecaption{ Objects in SIMBAD used to rule out possible evolved targets.\label{tab1}}
	\tablehead{
		\colhead{Object Type in SIMBAD} & \colhead{Object Type in SIMBAD}
	}
	%	\colnumbers
	\startdata
	Star            &  Nova \\
	OH/IR star      &  Post-AGB Star \\
	Carbon Star     &  Nova-like Star \\
	Variable Star   &  Mira candidate \\
	Wolf-Rayet Star &  Planetary Nebula \\
	Emission-line Star & AGB Star candidate \\
	Possible Carbon Star & Post-AGB Star Candidate \\
	Long-period variable star & Possible Planetary Nebula \\
	Variable Star of RR Lyr type & Variable Star of Mira Cet type \\
	Asymptotic Giant Branch (AGB) Star 
    \enddata
\end{deluxetable*}

\begin{figure*}
\centering
	\includegraphics[width=0.6\textwidth,height=0.4\textwidth]{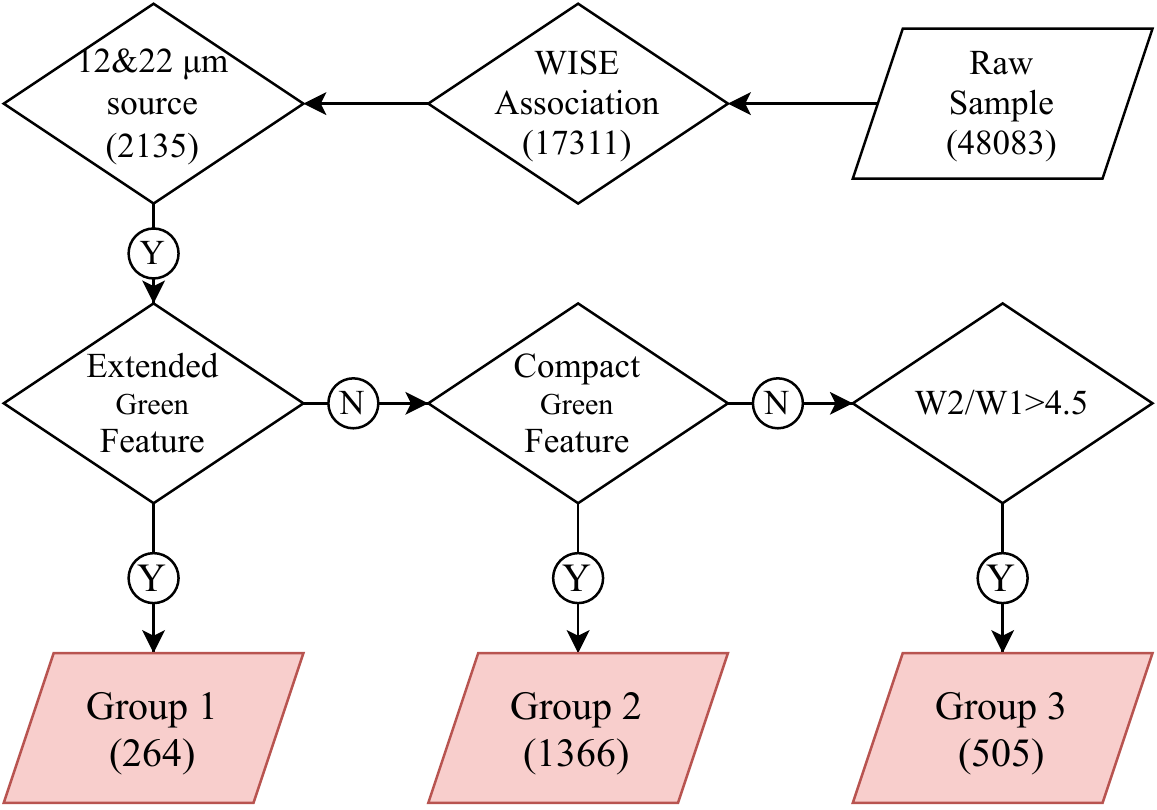}
	\caption{Flowchart describing the identification procedure of WGOs.}
\label{flowchart}	
\end{figure*}

%\citep{https://doi.org/10.26131/irsa1} 
We introduce detailed process of screening WGOs in this section. We first divide the W2 image by the W1 image pixel by pixel, pixels with an intensity ratio greater than 1.7 are retained, and those less than 1.7 are discarded, after that, sources with a size $\gtrsim$ one beam ($6\arcsec\times6\arcsec$) were extracted as ``Raw Sample". Then, we matched the ``Raw Sample" and ALLWISE source catalog \citep{https://doi.org/10.26131/irsa1} within 6\arcsec\ and obtained 17311 ``WISE Associations". YSOs with thick accretion disks usually radiate significantly in the 12 or 22 $\mu$m and can be detected even at substantial distances \citep{Wright2010}. So we further selected the samples with emission in the 12 or 22 $\mu$m to ensure that they are YSOs in an early accretion stage. In addition, based on our investigation, the ratio of W2 and W1 of most (over 80\%) identified EGOs in WISE is greater than 4.5. So according to this feature, if the ratio of the intensity of the 4.6 $\mu$m band to the intensity of the 3.4 $\mu$m band is more significant than 4.5, such a source we also temporarily consider as WGOs. WGOs cannot be directly identified by the ratio of 4.6 $\mu$m to 3.4 $\mu$m band alone, and this feature also appears in the late stages of some stellar evolution, such as asymptotic giant branch (AGB) star \citep{Busso1999, Herwig2005}, which makes us confuse them with WGOs. Therefore, we rule out possible targets in Table \ref{tab1} from WGOs candidates with non-green features by cross-identifications with the SIMBAD database in 6\arcsec ~(WISE approximate resolutions at 3.4, 4.6, and 12 $\mu$m), which contains valuable object type information of our targets \citep{2000A&AS..143....9W}. Fig.~\ref{flowchart} is a flow chart for identifying WGOs and the final number of the WGOs is 2135.

\begin{figure*}[ht!]
    \includegraphics[width=1\textwidth]{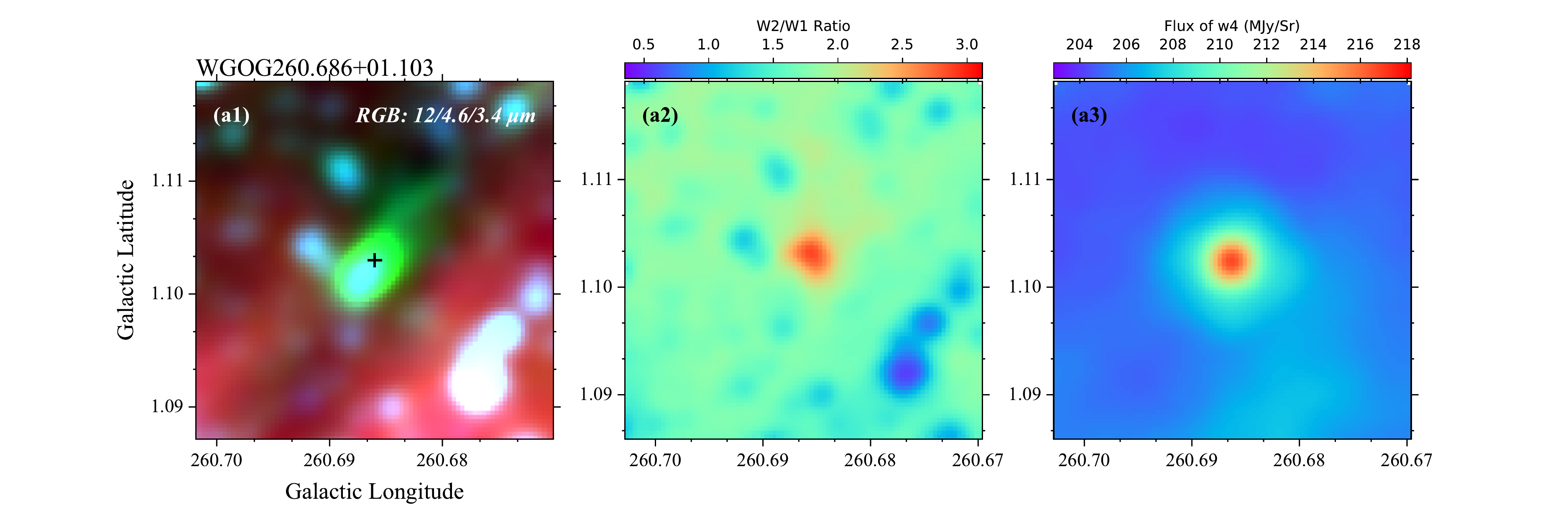}
	\includegraphics[width=1\textwidth]{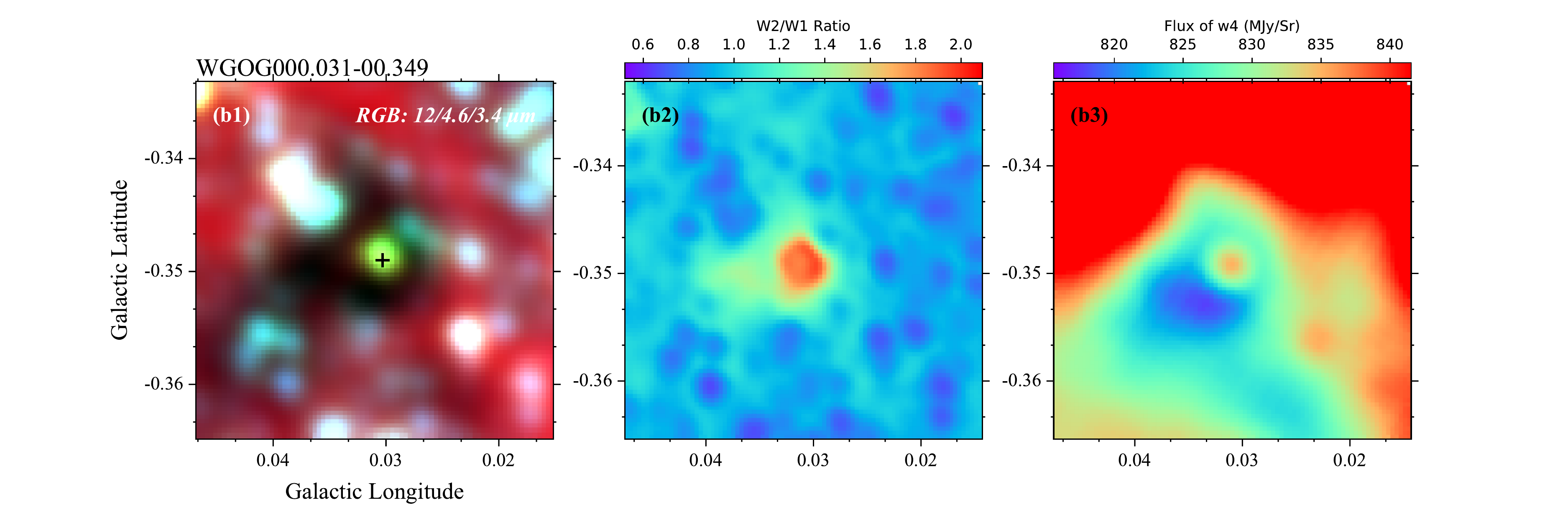}
	\includegraphics[width=1\textwidth]{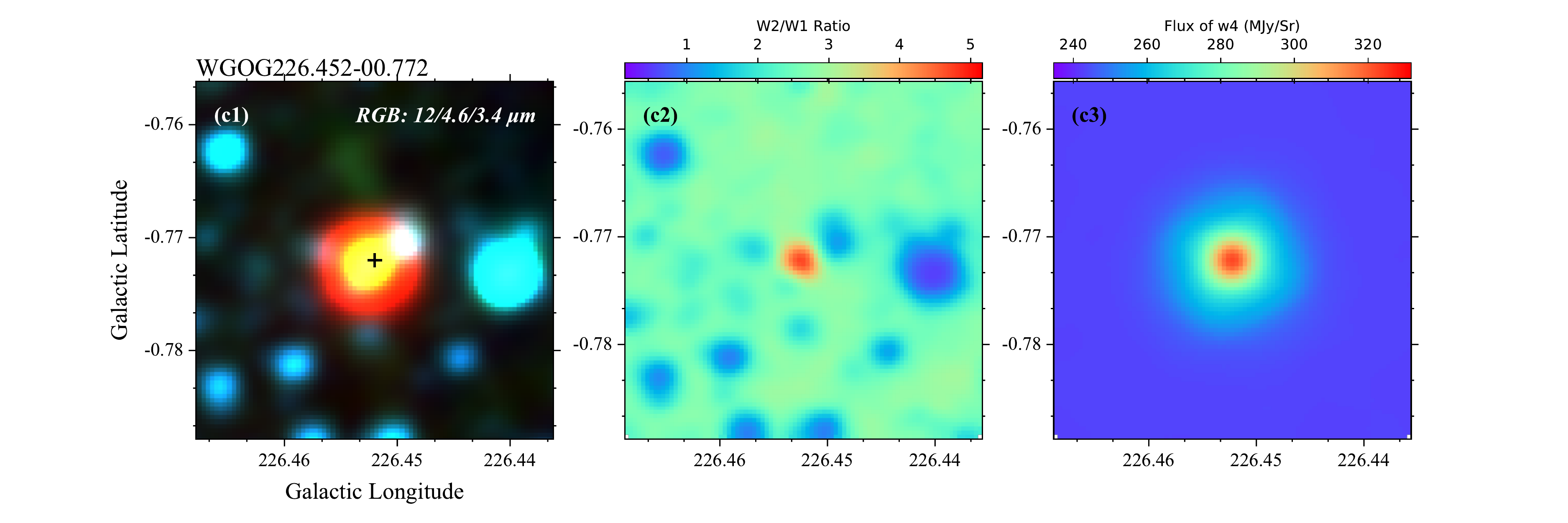}
	\caption{Morphology of the exemplary WGOs. Panels a, b, c represent groups 1, 2 and 3. Panels a1, b1, c1 are trichromatic maps of emission at 3.4, 4.6, 12 microns, rendered in blue, green, and red. Panels a2, b2, c2 are W2/W1 ratio maps. Panels a3, b3, c3 show 22 \micron~images. Black cross marks the center of each WGO.\label{fig:rgbRatio}}
\label{exemplary}		
\end{figure*}

To effectively identify the extended or compact green objects, we displayed WISE 3-color image into about 2.43 deg$^2$ (1.56\degree $\times$ 1.56\degree) mosaics, in which 3.4 $\mu$m (W1) band for blue channel, 4.6 $\mu$m (W2) band for green channel, 8.0 $\mu$m (W3) band for red channel. In addition, images of W2/W1 in the same format as the 3-color images were also made to search for WGOs that were not green. Each image was independently visually searched twice by the three authors of this paper, and only WGOs approved by all three were added to the catalog. All WGOs are divided into three sub-catalogs according to morphological features and the intensity ratio of W2 and W1 (Fig.~\ref{flowchart}).  Extended green structure is an obvious external
feature of MYSO. WGOs with extended green structures are classified as Group 1, with detailed selection criteria being that the ratio of the major axis to the minor axis of the source is $\gtrsim$ 1.2. WGOs with green compact morphology but without extended structure are classified into Group 2. Group 3 has no apparent green color feature, but its intensity ratio of W2 and W1 is more significant than 4.5. Fig.~\ref{exemplary} shows an example of the WGOs in Groups 1, 2, and 3. Since groups 2 and 3 do not yet have extended structures formed by outflow activity, they may be younger than traditional EGOs and may be MYSO candidates at an early stage. The number of groups 1, 2, and 3 are 264, 1366, and 505, accounting for 12\%, 64\%, and 24\% of the total.

%WISE green objects
%(a1) (b1) (c1) three color map with emission at 3.4, 4.6, 12 \micron~rendered to be blue, green, and red.

\begin{figure*}
  \centering
  \begin{minipage}[t]{1.0\linewidth}
  \centering
     \includegraphics[width=0.45\textwidth]{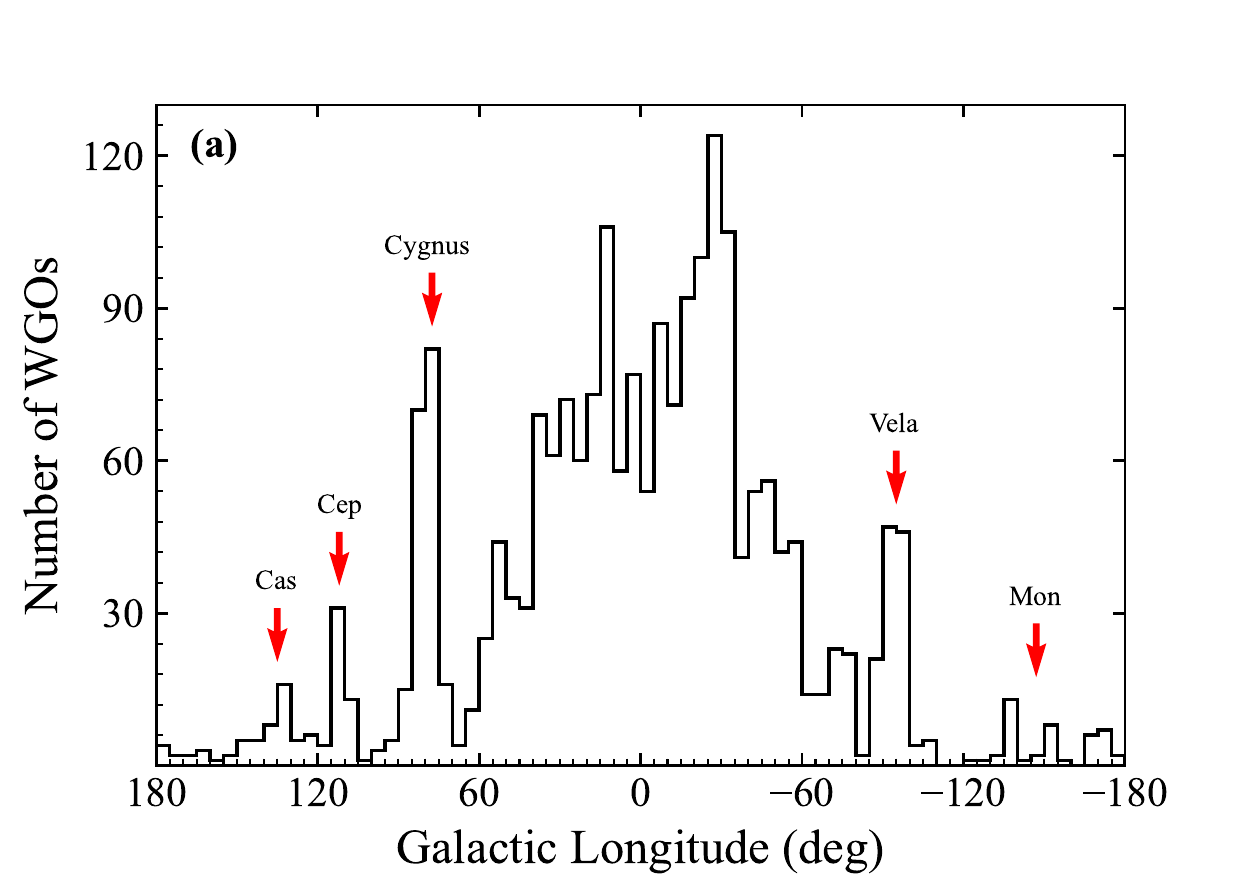}
     \includegraphics[width=0.45\textwidth]{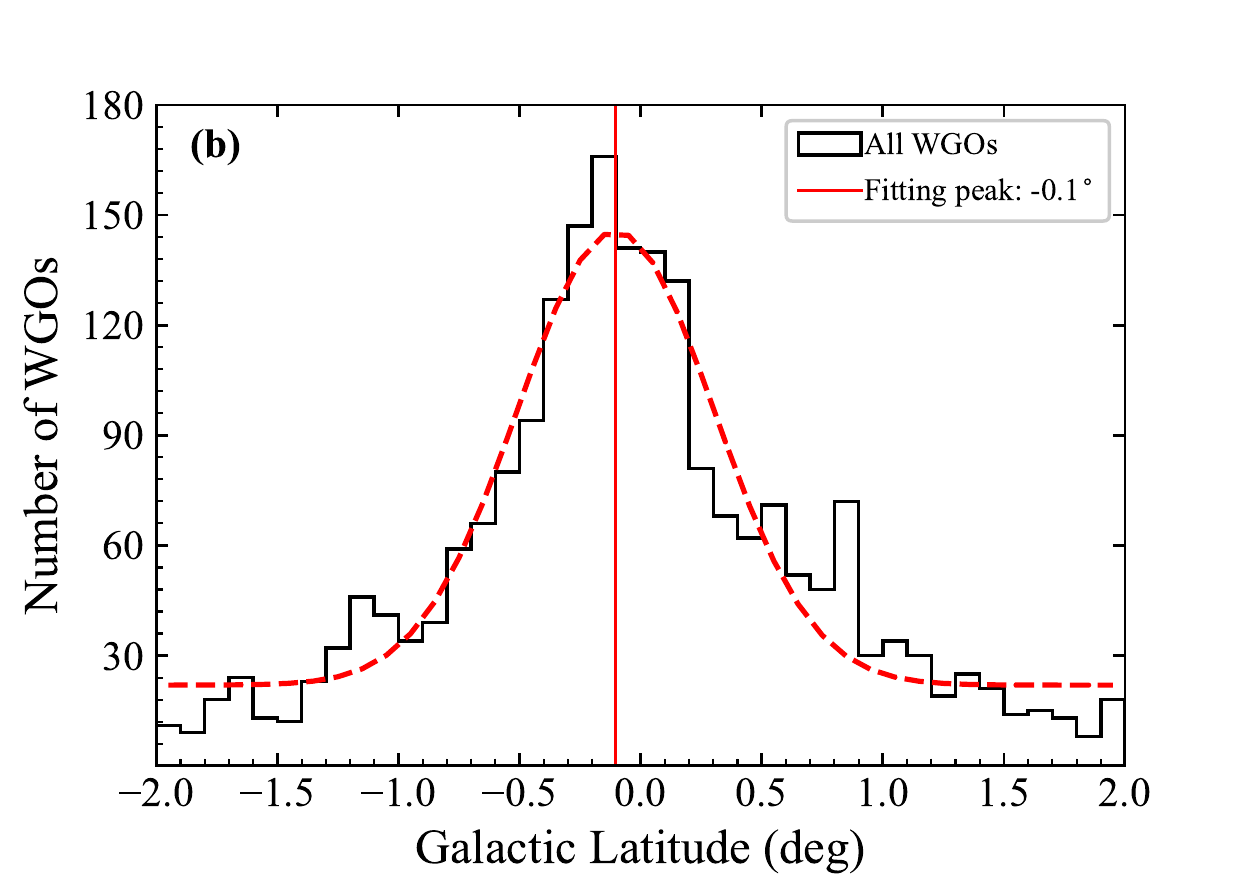}
  \end{minipage}  
\caption{
Panel (a) is the number distribution of WGOs as a function of Galactic longitude in 5\degree~bins. The locations of giant molecular clouds are indicated by red arrows.
 The panel (b) is number distribution of WGOs as a function of Galactic latitude in 0.1\degree~bins and fitted with a Gaussian curve. The red vertical line denotes the peak position of the Gaussian curve.}
\label{l_distribution}
\end{figure*}

\begin{figure*}
  \centering
     \includegraphics[width=0.85\textwidth]{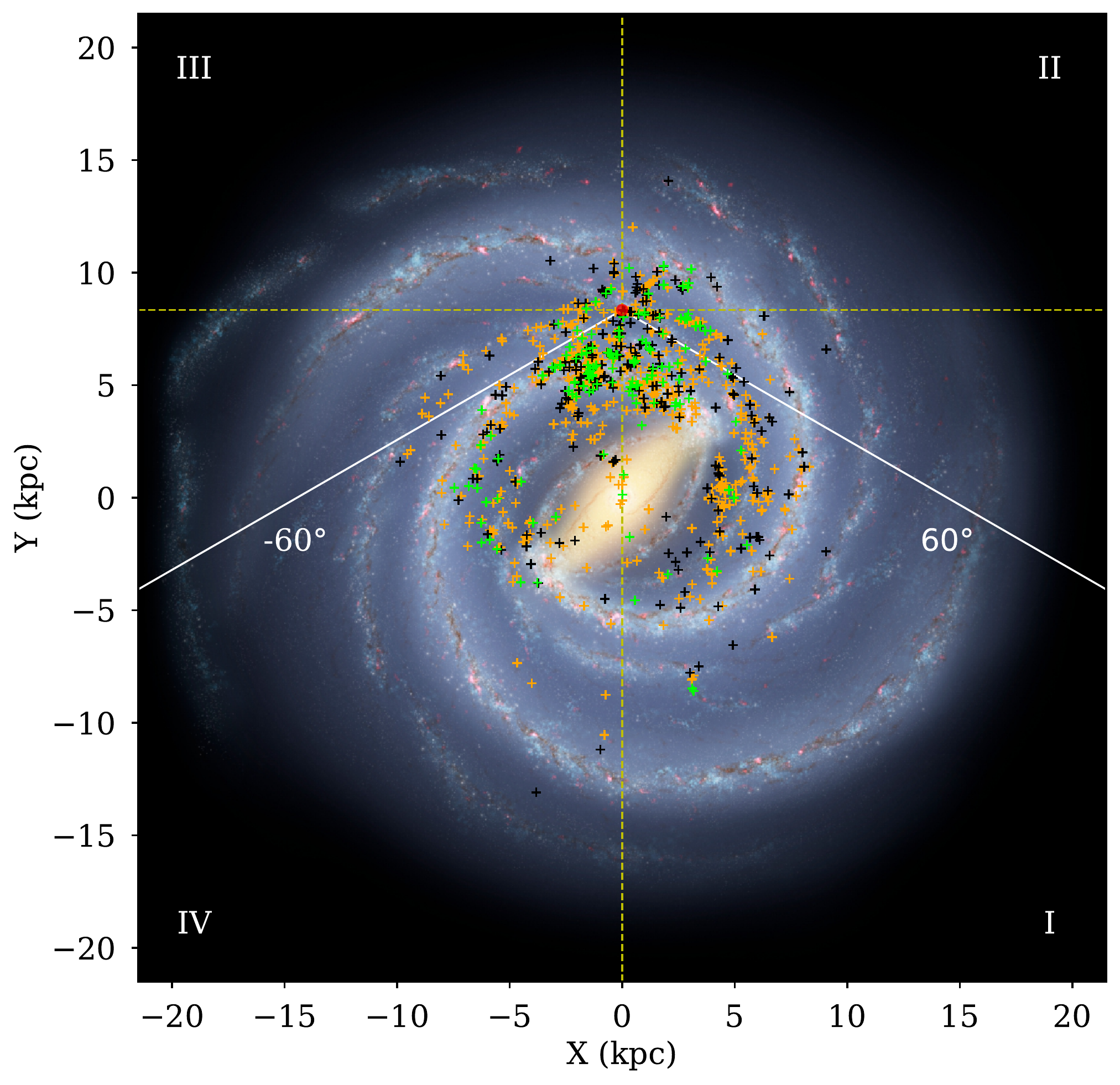}
\caption{The distribution of WGOs in the Galactocentric system. The axes are Galactocentric distances. We mark the WGOs of three different groups with green, yellow and black crosses, respectively. The background image is a sketch of the Galaxy produced by Robert Hurt (artist’s concept, R. Hurt: NASA/JPL- Caltech/SSC). The position of the Sun is shown by the red circle above the Galactic midplane. The two solid white lines enclose the inner Galactic Plane ($\mid l \mid<$ 60\degree).}
\label{scale_height}
\end{figure*}
\subsection{Galactic distribution of the WGOs} \label{sec:3.2}
2135 WGOs are shown in Fig.~\ref{l_distribution}a. 1578 WGOs are in $\mid l \mid <60$\degree, which make up 74\% of the total sample. WGOs in $\mid l \mid <60$\degree~are not symmetrically distributed. 709 WGOs locate in the range of [0\degree, 60\degree] with a peak at $\sim $15\degree. 869 WGOs locate in [-60\degree, 0\degree] with a peak at $\sim $25\degree. The number in [-60\degree, 0\degree] is 20\% more than that in [0\degree, 60\degree]. 556 WGOs are in $\mid l \mid >$ 60\degree. WGOs mainly distributed in 75\degree$ <l <$ 85\degree, 105\degree $ <l <$ 115\degree, 130\degree$ <l <$ 135\degree, -105\degree$ <l <$ -95\degree\ and -155\degree$ <l <-135$\degree\ regions, which are coincident with the locations of the Cygnus, Cepheus, Cassiopeia, Vela and Monoceros giant molecular clouds respectively. 

 As the parent structure of the YSO, the size and mass of the Hi-GAL compact source determine the potential to form the massive star. We cross-matched WGOs and Hi-GAL compact source catalog, and found 1260 WGOs have Hi-GAL source counterparts. Among them, 1151 sources have complete information from \cite{2021MNRAS.504.2742E}, such as heliocentric distance, luminosity, mass, and temperature. We adopt the distance of the Hi-GAL sources as the distance of the corresponding WGOs. Fig.~\ref{scale_height} gives an overview of the distribution of those 1151 WGOs in the Milky Way viewed from the north Galactic pole. If we only infer the distribution of WGOs from Fig.~\ref{l_distribution}, an erroneous conclusion will be obtained, that WGOs are concentrated in the Galactic Center, because the projection effect is not considered. But in Fig.~\ref{scale_height} we can see that most of the WGOs are distributed along the spiral arm. Many WGOs in the line-of-sight direction are just projected to the CMZ, but the actual distance is not that far. 

The red dashed line in Fig.~\ref{l_distribution}b shows the distribution of WGOs on the Galactic latitude ($\mid b \mid <2$\degree), which can be well fitted with a Gaussian curve, with FWHM of 1\degree~and a peak of $\sim $150. In the FWHM range ($\pm 1$\degree) of this Gaussian curve, the number of WGOs accounts for 80\% of the total sample. The center position of this Gaussian curve is not at the center of the galactic coordinates but shifted to the left by 0.1\degree~(see red vertical line in Fig.~\ref{l_distribution}b). This small offset also exists in the distribution of the core or clump structures such as the ATLASGAL survey \citep{2009A&A...504..415S} and the Hi-GAL survey, with negative peak of latitude ($b$ = -0.05\degree~and -0.09\degree~ respectively). This may be since our solar system is slightly above the Galactic Plane, and the viewing line of sight causes this offset \citep{2009A&A...504..415S}, but \cite{2009ApJS..181..227H} argues that there are more molecular clouds in positive longitude, obscuring the dense source that results in this offset.
\subsection{Cross-match with previous catalogs}
\begin{figure}
\centering
\includegraphics[width=1\textwidth]{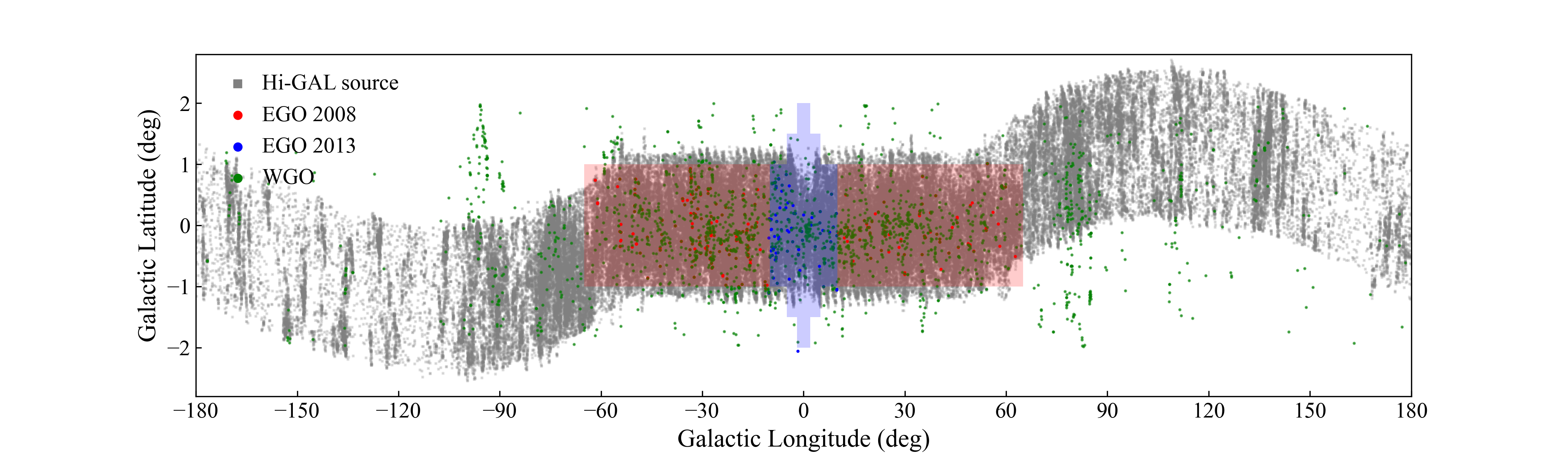}
\caption{Distribution of WGOs, EGOs and Hi-GAL sources in the galactic plane. Green circles represent WGOs; red and blue circles represent EGOs identified by \citet{Cyganowski2008} and \citet{Chen2013} respectively. The grey points show the Hi-GAL compact sources \citep{2021MNRAS.504.2742E}. The coverage of GLIMPSE I and II surveys are shaded in red and blue, respectively.}
\label{WGO_EGO}
\end{figure}

 Using {\it Spitzer} survery data, \citet{Cyganowski2008} cataloged $\sim $ 300 EGOs and \citet{Chen2013} cataloged 98 EGOs. There are 1401 WGOs in the GLIMPSE I and II survey region (see red and blue regions in Fig.~\ref{WGO_EGO}). Identification of EGO relies on visual inspection of objects in {\it Spitzer} 3-color images for the presence of green and extended structures. The identification methods of WGOs are similar to that of EGOs, both are visual inspection, but our adopted sample also include the compact green sources and the sources with excess emission W2/W1$\gtrsim$4.5. We cross-matched our WGOs to {\it Spitzer} EGOs within 6\arcsec\ which is equivalent to the resolution of WISE 3.4 and 4.6 $\mu$m bands, and found 70, 97 and 41 WGOs in each group are previously discovered EGOs. About half of EGOs do not have a corresponding WGO. After viewing blow-up images for each band of WISE of those unmatched EGOs, we found that the main reason why these EGOs were not picked from the WISE data was the lack of flux in the W3 or W4 bands, which accounted for more than 76\% of the EGOs, and the remaining EGOs whose W2/W1 ratio failed to exceed our threshold of 1.7. There is no case where the source blends with adjacent targets due to the low resolution of the WISE and is not recognized.

 Due to the constraint of the (approximately $\mid b \mid<$ 1.5\degree) survey range of Hi-GAL \citep{Molinari2010}, 363 WGOs are distributed outside this range, and about 71\% (1260/1772) of WGOs have Hi-GAL counterparts within this range. By observing the positions of the remaining nearly 30\% WGOs in survey range (see grey points in Fig.~\ref{WGO_EGO}), we found that some of them located at the edge of the observation field and therefore cannot be effectively identified as sources. However, due to lack of dust emission, the possibility of a few objects in WGOs unrelated to star formation cannot be ruled out. This needs to be further confirmed by follow-up observations. About 36\% of the Hi-GAL sources are less than 0.1 pc in size which is the core scale, and the sizes of the remaining 64\% are between 0.1 and 1 pc at the clump scale. The median values of the luminosity, mass, and distance are 771 $L_{\odot}$, 320.6 $M_{\odot}$, and 3.7 kpc, respectively. We also cross-matched ATLASGAL YSOs catalog \citep{Urquhart2022} in inner Galactic Plane ($\mid l \mid <$ 60\degree) without $\mid l \mid<$ 3\degree, and we found that 1176 WGOs cannot match ATLASGAL YSOs.
 
\cite{2021ApJS..254...33K} presented a catalog of $\sim 1.2\times10^{5}$ YSOs based on the $\it Spitzer/IRAS$ survey (the inner galactic plane, approximately -105\degree$<l<$ 110\degree,~ $\mid b \mid<$ 2\degree). 1992 WGOs locate at this survey coverage area, and 1095 WGOs match YSOs within 6\arcsec. The match rate is 54.9\%. \cite{2016MNRAS.458.3479M} identified over 13300 YSOs using 2MASS and WISE photometric data combined with support vector machine in all sky. Our WGOs show a low match ratio with this catalog, with only 68 WGOs can match. Meanwhile, the Red {\it MSX} Source survey (RMS) \citep{2013ApJS..208...11L} provided a catalog with nearly 2800 YSOs, which included 115 MYSOs. 1859 WGOs locate in the range of RMS catalog (10\degree $<l<$ 350\degree), and we found that 1653 WGOs are newly identified.

\subsection{Cross-match with masers of the star formation indicators}
\label{sec:3.3}
\begin{figure*}
\centering	\includegraphics[width=0.75\textwidth]{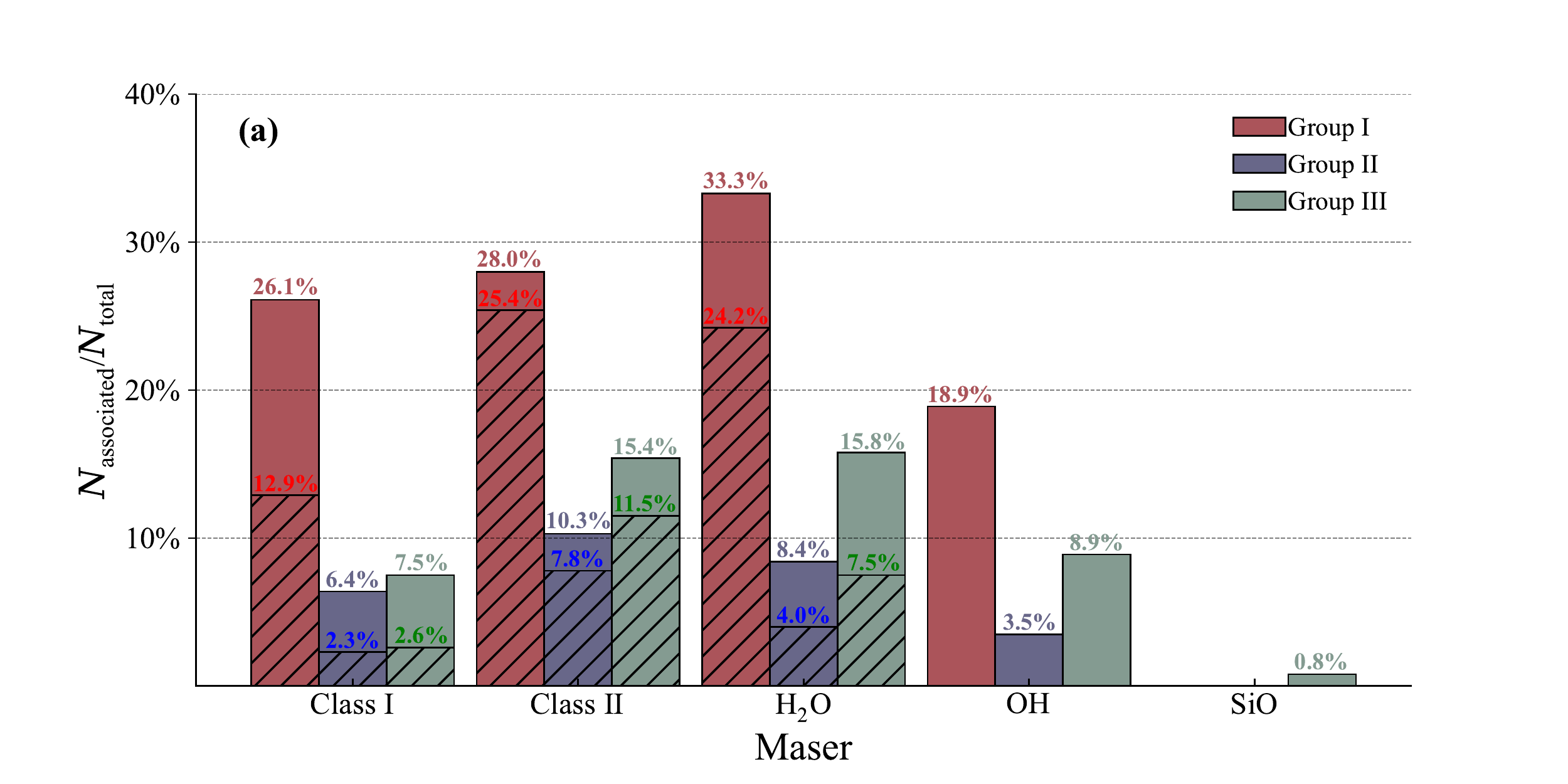}	\includegraphics[width=0.75\textwidth]{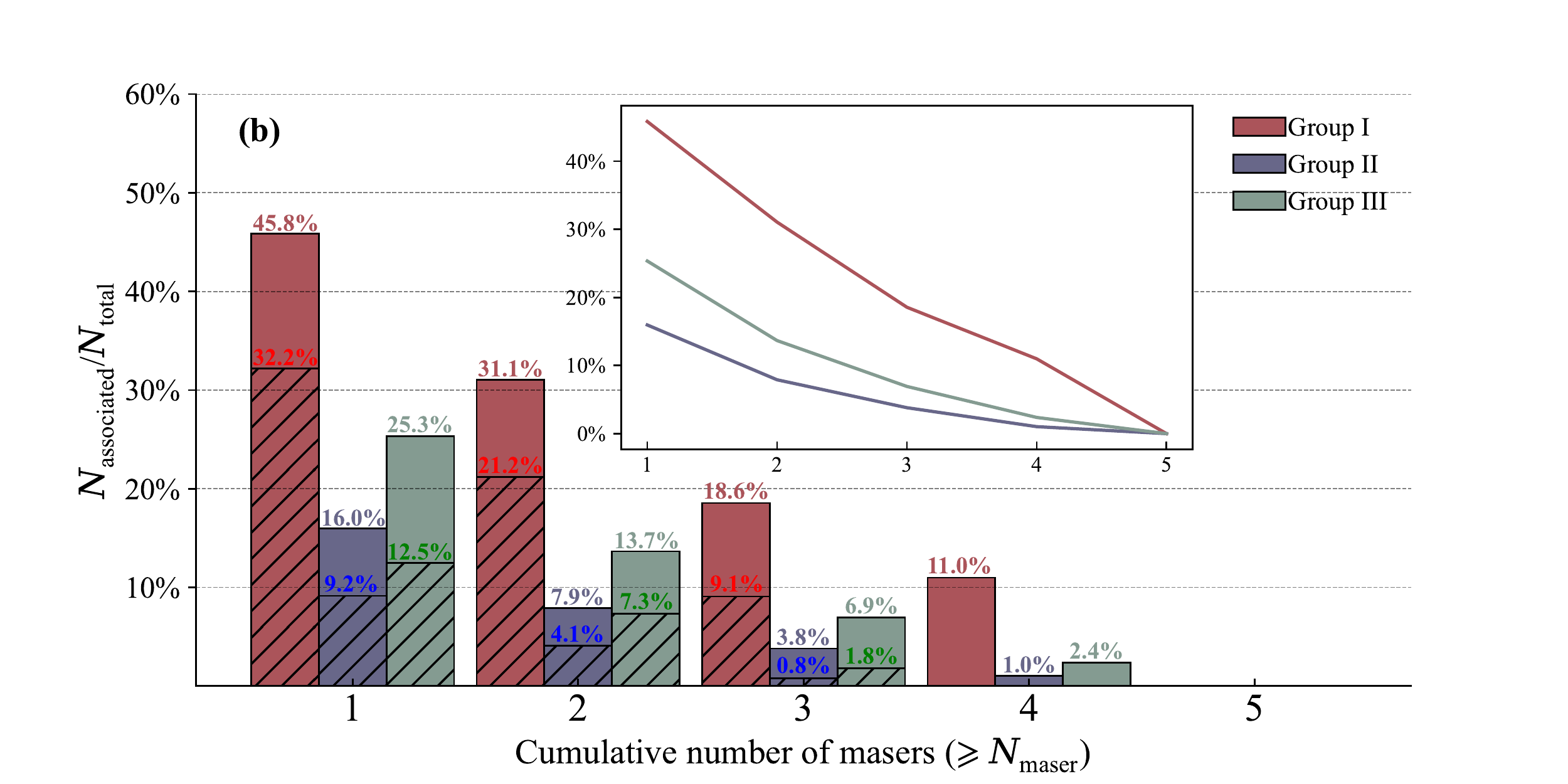}
\caption{The association rate of the WGOs in each group with five kinds of masers (methanol Class I/II, H$_{2}$O, OH, and SiO) (upper panel). $N_{\rm associated}$ is the number of the WGOs associated with the maser. $N_{\rm total}$ is the total number of the WGOs in each group. The red, blue and green bars represent Group 1, 2 and 3 respectively. The lower panel is the association rate of WGOs with the cumulative number of masers ($\geqslant N_{\rm maser}$). $N_{\rm associated}$ in this panel is number of WGOs associated with at least $N_{\rm maser}$ masers. The sub-panel in panel (b) shows the change of association rate between each group of WGOs and cumulative number of masers.}
\label{maser}	
\end{figure*}

In the early stage of star formation, MYSOs are deeply embedded within dense molecular cores, with a short accretion period, and remain in the core for a long time even after entering the main sequence phase. Fortunately, large-scale or unique molecular emissions triggered by massive star formation, such as masers, can help identify them \citep{Beuther2007}. We cross-matched WGOs with five masers ($\rm{CH}_{3}\rm{OH}$ Class I/II, $\rm H_{2} \rm O$, SiO and OH) related to star formation, using online tool, {\it MaserDB}\footnote{\url{https://maserdb.net/}}, provided by \cite{2019AJ....158..233L}. Since different distances of WGOs will have a considerable impact on the actual matching radius, and the information on them is incomplete, so we use a uniform matching angular radius that is WISE resolution of 6\arcsec, which is corresponding to $\sim$ 0.15 pc at WGOs mean distance of 5.17 kpc and close to a typical core size (0.1 pc, \citealt{Zhang2022}). Masers detected by single-dish and interferometric observations have been selected separately. We separately counted the association rate of the WGOs in each group with five kinds of masers and the association rate of WGOs with the cumulative number of masers, which are shown in Fig.~\ref{maser}a and b, respectively. {\it MaserDB} distinguishes the interferometric positions of methanol Class I/II, $\rm H_{2} \rm O$ masers from their single-dish positions, but no interferometric positions for OH and SiO masers are provided. Each bar in Fig.~\ref{maser} shows the association rate of all maser data, including interferometric and single-dish data, and the slash covered portion of each bar shows associated rate of only interferometric maser data. In Fig.\ \ref{maser}a, the shape of bars in every kinds of maser, except SiO, are similar, whether it is all maser data or only interferometric data. The most ubiquitous type of masers is water, as it is associated with different astronomical objects \citep{2003ApJS..144...71F,2005A&A...434..613S}, at different stages of evolution (e.g., protostellar jets, \citealt{2013ApJ...773...70H}; large-scale shocks, \citealt{1994ApJ...427..914M}; disks, \citealt{2003ApJ...586..306G}). It is also consistent with the statistical results shown in Fig.\ \ref{maser}a. While methanol masers are less ubiquitous. The sub-panel in Fig.\ \ref{maser}b more intuitively shows the change of association rate between each group of WGOs and cumulative number of masers. Statistics based on Fig.\ \ref{maser} shows that maserassociation rates in Group 1 are at least 3 and 2 times greater than Groups 2 and 3, and Group 3 is slightly higher than Group 2. %The statistical results in Fig.~\ref{maser}b also support the conclusions drawn in Fig.~\ref{maser}a. The association rate of Group 2 and Group 3 decreased rapidly with the cumulative number of masers, especially the rate of Group 2.
%and the distances of WGOs are different 

\subsection{Derived physical parameters of the WGOs from the SED models}
\begin{figure}
\centering
\includegraphics[width=0.6\textwidth]{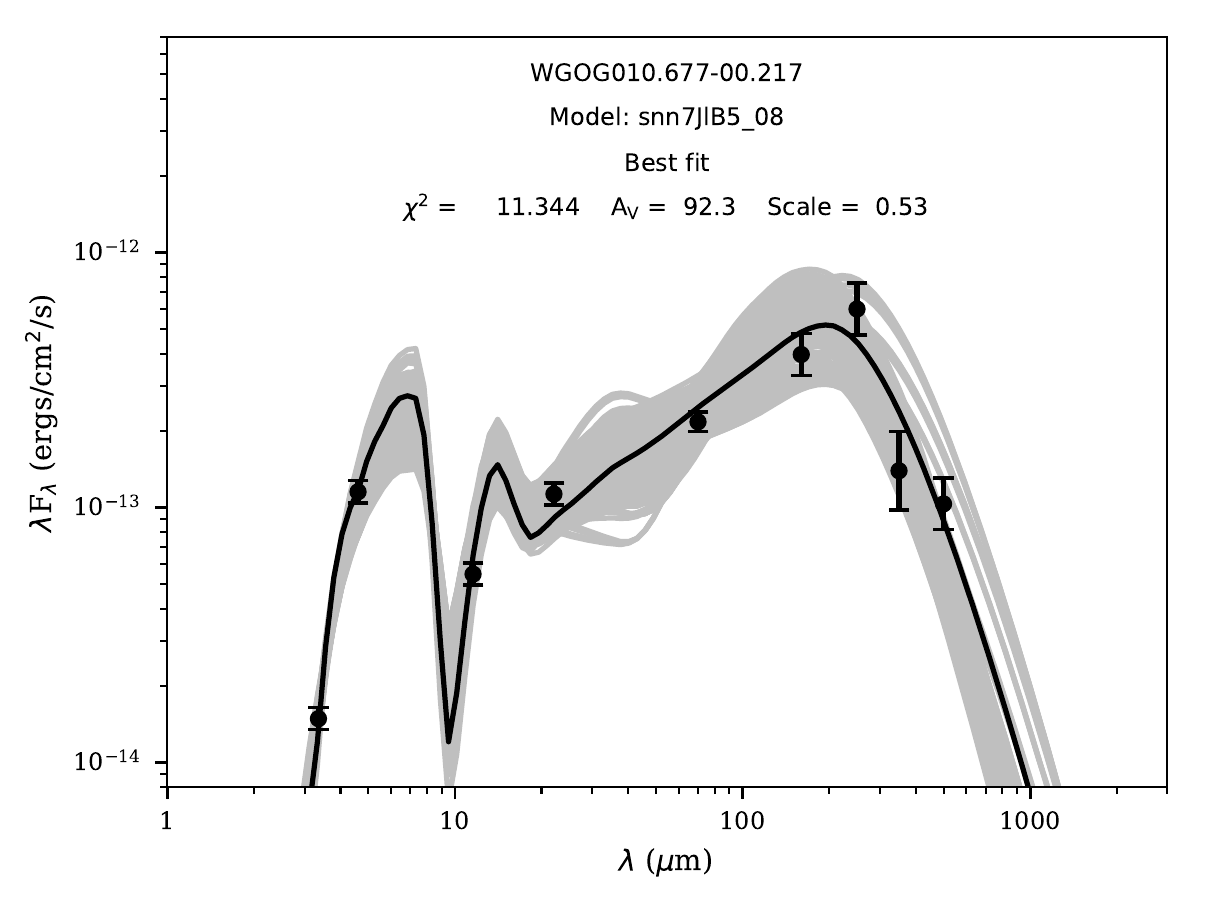}
\caption{An example of SED fitting. Photometric data from WISE and Hi-GAL were fitted using YSO models of \cite{Robitaille2017}. Black and gray lines represent best fit and good fit which satisfy $\chi^{2} - \chi_{\rm best}^{2}<3N_{\rm data}$, where $N_{\rm data}$ is the numbers of data points. The $P(D\mid M)$ score of this fitting is 4.564.
\label{SED}}
\end{figure}
%The intense material movement leads to the enhanced emission at the W2 band.
%It should be noted that water maser and methanol Class II maser have the highest association rate of all data and interferometric data respectively. 26.6\%, 7.2\% and 11.9\% are the mean association rate of each group. 

We constructed SEDs with WISE four and {\it Herschel} five wavelengths ranging from 3.4 to 500 $\mu$m for 1151 WGOs with Hi-GAL source counterparts. \cite{Robitaille2017} provided 18 kind of model sets for each contain 10,000 models for fitting YSOs in different evolution stage. Based on the prior knowledge of WGOs, there are maybe embedded protostars with an accretion disk or outflow. The model sets we selected are s-smi, sp-smi, sp-hmi, s-p-smi, s-p-hmi, s-pbsmi, s-pbhmi, s-u-smi, s-u-hmi, s-ubsmi, s-ubhmi, spu-smi, and spu-hmi. All 13 model sets contain one or more features, such as envelope or accretion disk or outflow. Fitting the SED of each WGO to the 13 different model sets return $\chi^{2}$ for each model set. We used $\chi^{2}-\chi_{\rm best}^{2}<3N_{\rm data}$ as a criterion to select good fits from all model sets. \cite{Robitaille2017} suggested using a Bayesian approach to compare how well among different model sets explain a set of data. $P(D\mid M)\propto N_{\rm good}/N$, where $N_{\rm good}$ is the number of good models from a given model set, and $N$ is the total number of models in that set, is used to indicate the reliability of the model set to the data. In this way, for each WGO, we only need to compare 13 different values of $P(D\mid M)$, and the highest value of $P(D\mid M)$ represents the most suitable model set. We show an example of SED fitting in Fig.~\ref{SED}. Based on the fitting results of the best model set for each WGO, we calculated the mean stellar radius $R_{\star}$ and surface temperature $T_{\rm eff}$, weighted by $1/\chi^{2}$. The uncertainties of $R_{\star}$ and $T_{\rm eff}$ are from weighted standard deviation (e.g., \citealt{2019ApJ...875..135T,2021ApJ...922...90C}).

The YSO total luminosity is composed of photospheric luminosity ($L_{\star}$) and accretion luminosity ($L_{\rm acc}$) (e.g., \citealt{2021ApJ...922...90C,2021ApJ...912L..17H,2021ApJ...908...68O})
\begin{normalsize} 
\begin{equation} 
\centerline{$L_{\rm tot}$ = $L_{\star}$ + $L_{\rm acc}$}.
\end{equation} 
\end{normalsize}
Here we assume all YSOs are absolute blackbody and spherical. According to the Stefan-Boltzmann law, the stellar luminosity $L_{\star}$ can be derived.

The median values of $L_{\star}$ in each group of WGOs are 16.6 $L_{\odot}$, 5.4 $L_{\odot}$ and 7.4 $L_{\odot}$ respectively. The median value of $L_{\star}$ in Group 1 is much larger, perhaps indicating that WGOs in Group 1 are more mature than the other two groups. The accretion rate $\dot M_{\rm acc}$ can be estimated from accretion luminosity \citep{2011ApJ...733...55Z}
\begin{normalsize} 
\begin{equation} 
\centerline{$L_{\rm acc}$ = $\displaystyle{\frac{GM_{\star}\dot M_{\rm acc}}{R_{\star}}}$},
\end{equation} 
\end{normalsize}
where $G$ and $M_{\star}$ are the gravitational constant and mass of protostar. $M_{\star}$ is calculated according to the method proposed by \cite{1996MNRAS.281..257T}. The mean $M_{\star}$ of each group are 2.6$\pm $1.6 $M_{\odot}$, 1.9$\pm $0.9 $M_{\odot}$, and 2.3$\pm $1.5 $M_{\odot}$ respectively. The Hi-GAL source is the parent structure of the protostars, which wraps the protostars, and its luminosity can be regarded as the total luminosity of the protostars. It should be noted that protostars are embedded in cores or clusters, and multiple protostars may be accreting simultaneously \citep{2009MNRAS.400.1775S, 2017MNRAS.468.3694C}, but massive protostars are usually much more luminous than low-mass protostars. On the contrary low-mass protostars contribute little for the luminosity. We obtained the median values of $\dot M_{\rm acc}$ of each group are $2.04 \times 10^{-4}\ M_{\odot} \rm yr^{-1}$, $2.91 \times 10^{-5}\ M_{\odot} \rm yr^{-1}$, and $4.22 \times 10^{-5}\ M_{\odot} \rm yr^{-1}$, respectively. 

\section{DISCUSSION} \label{DISCUSSION}
\subsection{The star formation scenario of the WGOs}
Whether a source can be successfully detected depends on its intensity in the direction of sight. 4.5 $\mu$m emission of the EGOs identified by \citet{Cyganowski2008} is $\gtrsim 4~{\rm MJy~sr^{-1}}$, and the flux intensity of most of the EGOs in {\it Spitzer} 4.5 $\mu$m band is significantly higher than that in 3.6 $\mu$m band \citep{2004ApJS..154..352N,Cyganowski2008}. 
We screened 2135 WGOs as young candidates for massive stars, significantly outnumbering {\it Spitzer} EGOs, which benefits from WISE being twice as sensitive as {\it Spitzer} \citep{2003PASP..115..953B,Wright2010}, even though they are filtered by multiple thresholds we set. The empirical relationship between the star formation rate (SFR) in our galaxy and the physical properties of interstellar gas is known as ``Kennicutt–Schmidt law": $\Sigma_{\rm{SFR}}$ $\propto$ $\Sigma_{\rm{gas}}^{N}$. This relation states that the SFR is positively related to gas density with the index of the power-law being $\sim $2 in the solar neighborhood \citep{Schmidt1959} and 1.4 in a larger set of galaxies \citep{Kennicutt2012}. Cold dust view reveals the confined dust lane in the Galactic Plane, which is bright in -48\degree\ $<l<$ 40\degree, but outside this range gets weaker significantly \citep{Csengeri2016}. WGOs are mainly distributed in $\mid l \mid <$ 60\degree, which are synergistic with the gas in the plane of the Milky Way. The central molecular zone (CMZ, the inner $\pm$1.5\degree\ $\times$ $\pm$0.5\degree~around the Galactic center) contains $\sim 10$\% of neutral gas of Galaxy ($\sim5 - 10 \times 10^{7}\ M_{\odot}$) \citep{2017IAUS..322..147B}, but only accounts for 0.1\% of the surface area. CMZ has a large amount of dense molecular gas, however due to the extreme environment in this region, its star formation efficiency is very low, only $\sim 0.07 - 0.15\ M_{\odot}\rm yr^{-1}$ \citep{2011MNRAS.413..763C,2009ApJ...702..178Y,2011ApJ...736..133A}. At the same time, it lacks YSOs with high mass and low mass (total mass of YSOs in the CMZ are $\sim 7.7 \times 10^{4}\ M_{\odot}$) \citep{2012A&A...537A.121I}. The number of our WGO samples is lack in the CMZ, which is consistent with the low SFR in this region.

According to the classification method of the WGOs, Group 1 WGOs are most in line with the morphological characteristics of traditional EGOs \citep{Cyganowski2008,Chen2013}. The Group 2 WGOs lack the extended structures but show compact green features. The Group 3 WGOs have neither extended structure nor green features, but W2 is 4.5 times larger than W1. The resolution of WISE is about three times lower than that of {\it Spitzer}, so extended structures of some WGOs may be not identified and are classified into the Group 2 or 3. WGOs in the Groups 2 and 3 do not exhibit extended structures, they show relatively strong W2 band emissions and are usually surrounded by dense gas, and they may be in an earlier stage (e.g., Class I even Class 0 of YSO, \citealt{1996ARA&A..34..111B,Shu1987}) than the Group 1 and are currently accreting gas. 

%Therefore, the association rate with the molecular tracers can infer the evolution sequence of YSOs to a certain extent.
Molecular tracers behave differently through the evolutionary sequence, and the late evolutionary sequence may contain more complex molecular tracers \citep{2012ApJ...756...60S}. Different masers usually indicate different evolution stages of YSOs, and the star-forming regions with later evolution stages show more complex maser components (e.g., \citealt{2011A&A...527A..32W}).  Class I methanol masers excited by the shock waves are usually found at some distance from a radiation source \citep{sobolev2007}. It is worth noting that it may occur at the earliest stage, earlier than all masers \citep{1992MNRAS.259..203C,2006ApJ...638..241E}. Class II methanol masers are widely considered to be one of the most reliable tracers in the early stage of high-mass star formation (\citealt{2006ApJ...638..241E}, and references
therein), and this maser are only related to high-mass star formation activities \citep{2013MNRAS.435..524B}.
We speculate that the WGOs of Group 1 are in the latest evolutionary stage compared to the other two groups, and the shock wave excited by the strong star formation activity that even excited them near the radiation source leads to the highest association rate with Class I maser near the radiation source. The inference about the evolution stage of Group 1 can also be deduced by observing the association rate of the OH maser: the OH maser is a sensitive tracer of Ultra-Compact \HII (UC \HII) region \citep{2002IAUS..206..506R}, and the occurred of UC \HII region indicates a later stage of star formation. Several previous studies have shown that water maser in star-forming regions can be excited by a variety of star-forming activities (e.g., \citealt{2013ApJ...773...70H, 1994ApJ...427..914M, Gallimore_2003}), and it is obvious that various activities occurring with the process of evolution will improve the excitation rate of water master. The SiO maser have been known associated with late-type stars, such as stars on the AGB \citep{2000PASJ...52..895M,2000PASJ...52L..43N}. The SiO maser were confirmed that it in star-forming regions is a rare phenomenon by \cite{2009ApJ...691..332Z} as it was only detected from known regions (e.g., Orion KL, \citealt{1974ApJ...189L..31S}; W51 North, Sgr B2, \citealt{1986mmmo.conf..275H};  Sgr B2(N), \citealt{2015ApJ...815..106H}; G19.61-0.23 and G75.78+0.34, \citealt{2016ApJ...826..157C}). This also leads to the fact that in Fig.\ \ref{maser}a, SiO maser appears to be almost unassociated with any WGOs.

\subsection{How many WGOs are high reliable MYSOs?}
The extended structure of the EGOs is likely to be related to the molecular outflow resulting from the shock emission from the molecules H$_{2}$ and CO during the formation of a massive star \citep{Cyganowski2008,Chen2013}. The WISE 4.6 $\mu $m is very close to Spitzer’s 4.5 $\mu$m, and we believe that the extended structure traced by WISE 4.6 $\mu $m is also due to the outflow. Outflow is observed toward both low-mass (e.g., famous well researched Herbig–Haro object HH46/47, \citealt{2004ApJS..154..352N,2007ApJ...668L.159V}) and high-mass sources. Strong outflow is one feature of MYSOs. Compared to the flux rate of low-mass outflow $\sim 10^{-6}M_{\odot}{\rm yr^{-1}}$, the flux rate of MYSOs can be stronger than $10^{-3}M_{\odot}{\rm yr^{-1}}$ \citep{2007prpl.conf..245A}. The bright emission in the W2, which maybe a precursor to a strong outflow in future evolution. However, it cannot be ruled out that there is not enough gas in the parent structure to provide enough gas for it to evolve into a massive star. Thanks to the multi-band SED models for YSOs, we can predict the evolution of the sources and provide us with a way to get more believable subsamples \citep{2006ApJS..167..256R, Robitaille2017}. Massive stars form in clusters \citep{2003ARA&A..41...57L,2012MNRAS.426.3008K}, there may be multiple YSOs for one WGO (e.g., MM1$\sim$19 in EGO11.92-0.61, \citealt{2017MNRAS.468.3694C}). Limited by the resolution of the WISE, it's indistinguishable. But the massive YSOs have a strong gravitational advantage that will limit the mass of other YSOs in the cluster, once them formed, would dissipate the natal cloud, preventing further star formation (e.g., \citealt{1962AdA&A...1...47H}). Meanwhile, the luminosity and accretion rate of the massive YSOs are far higher than those of other low-mass YSOs \citep{2007prpl.conf..245A,Hosokawa2010}.

\begin{figure}
\centering
\includegraphics[width=0.7\textwidth]{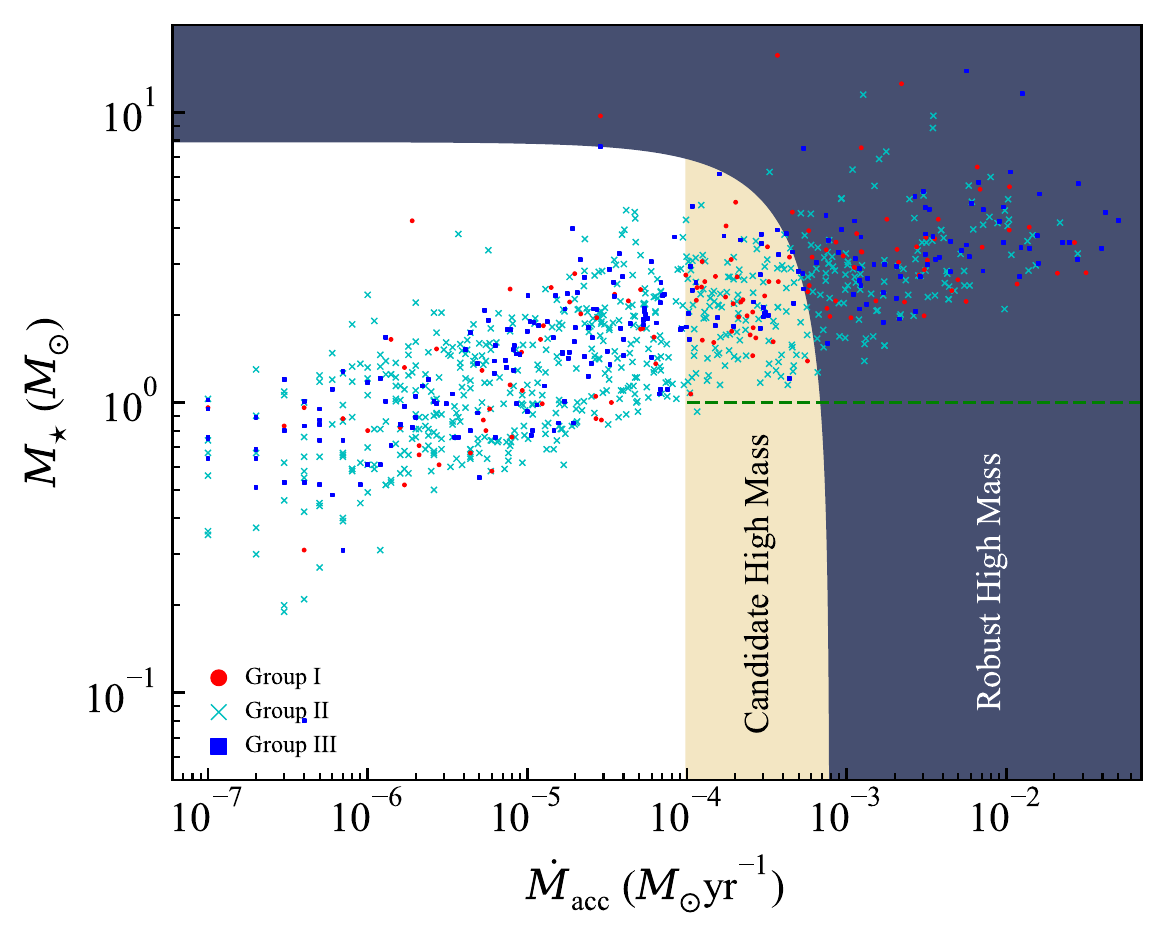}
\caption{Accretion rate versus protostellar mass. Red circles, blue squares, and cyan crosses represent WGOs in groups 1, 2, and 3, respectively. The WGOs listed in table \ref{tab2} and \ref{tab3} are located in the deep blue and light yellow areas, respectively. 
The green dotted line marks the mass equal to 1 $M_{\odot}$, and almost all MYSO candidates are above this line.
\label{M_acc}}
\end{figure}

\begin{figure*}
\centering
\includegraphics[width=0.75\textwidth]{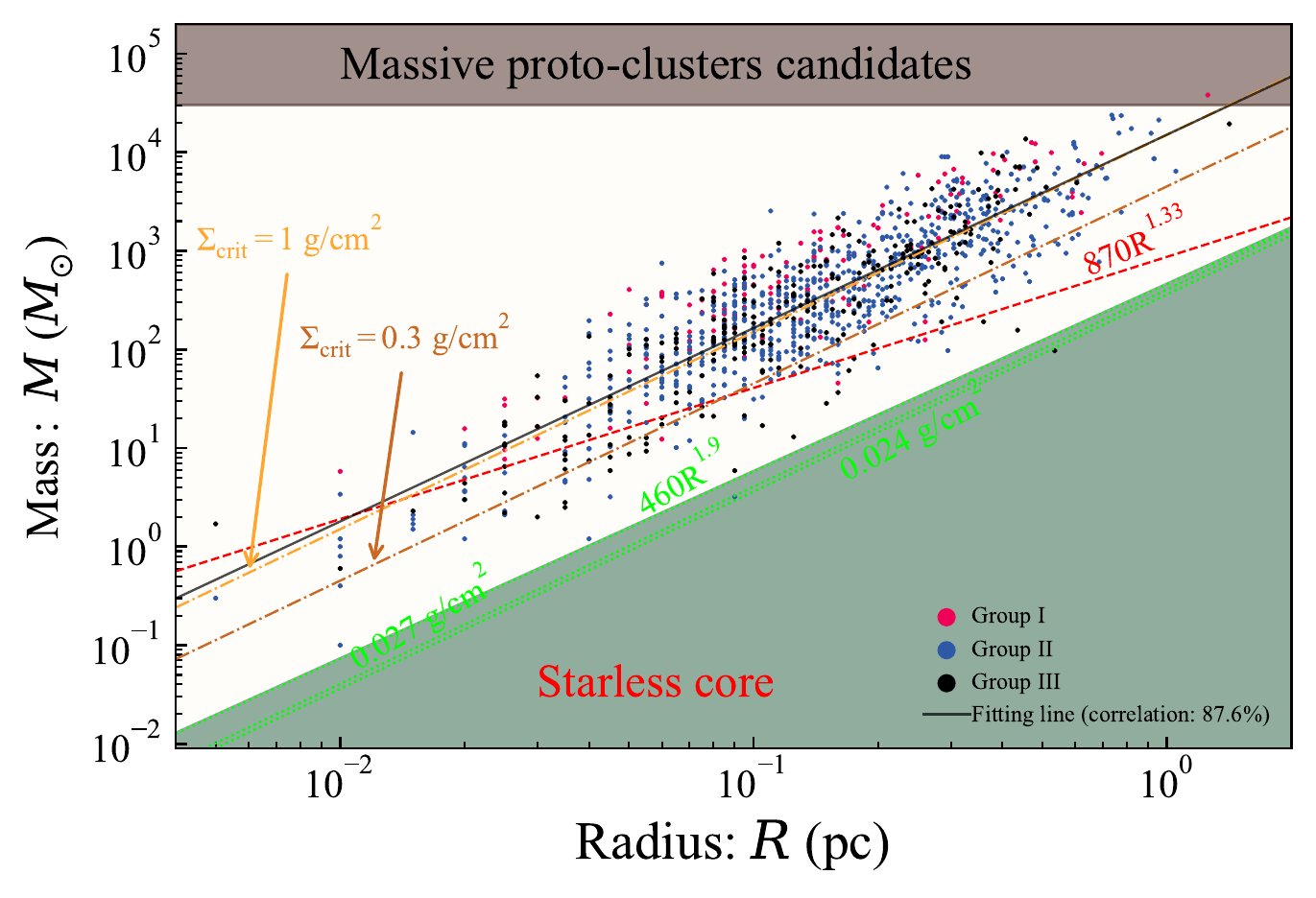}
\caption{Mass and radius diagram of Hi-GAL compact sources corresponding to WGOs in each group. The black solid line shows a clear correlation between source radius and mass, ${\rm log}_{10}(M/M_{\odot })=1.96{\rm log}_{10}(R/{\rm pc})+4.18$. The red dashed line shows the sources which may form high mass star. The yellow dash-dotted indicates another higher threshold ($\Sigma_{\rm crit}=1 \rm g/cm^{2}$) for judging high mass clumps. The deep yellow dash-dotted line shows lower threshold than the yellow one.}
\label{R_M}
\end{figure*}

%Only few WGOs' Hi-GAL source counterparts are under this line
We selected the WGOs with $M_{\star}+\dot M_{\rm acc} \times 10^{4}\ {\rm yr}>8 \ M_{\odot}$ (see  blue area in Fig.~\ref{M_acc}) as robust MYSO candidates, which are listed in Table \ref{tab2} and the numbers in each group are 46, 118, and 67. 
The accretion rate changes dramatically during massive star formation, but considering the accretion time scale is usually shorter than $10^{4}$ years, it is generally believed that the accretion rate of MYSOs is $\sim 10^{-4}-10^{-3}\ M_{\odot}\rm{yr}^{-1}$ \citep{2011ApJ...733...55Z}. 
We selected WGOs with $\dot M_{\rm acc}>10^{-4}\ M_{\odot}\rm{yr}^{-1}$ (see Fig.~\ref{M_acc} yellow area)  as candidate MYSOs. Table \ref{tab3} lists those candidate MYSOs. The numbers in each group are 32, 110, and 30. 63\% WGOs in Group 1 are robust or candidate MYSOs, much higher than the ratio in Group 2 and 3 (34\% and 40\%). This probably implies that the WGOs Group 1 have the greatest potential to form massive stars. By observing Fig.~\ref{M_acc}, we found that almost all stellar mass of MYSO candidates are higher than 1 $M_{\odot}$.
Based on the relation of masses and radii of the Hi-GAL sources, we inferred whether massive stars will form. Prestellar and starless cores are separated by the three dotted green lines $\Sigma_{\rm crit} = 0.024~\rm g~\rm cm^{-2}$ \citep{2010ApJ...724..687L}, $\Sigma_{\rm crit} = 0.027~\rm g~\rm cm^{-2}$ \citep{2010ApJ...723.1019H} and $M(r) = 460M_{\odot}(r/\rm{pc})^{1.9}$ \citep{1981MNRAS.194..809L} in Fig.~\ref{R_M}, and the green area means the core in which is not bound by gravity may disperse after a period of time, so the probability of forming stars is very low. Only few WGOs' Hi-GAL source counterparts are under this line, therefore, we can infer that most WGOs' Hi-GAL source counterparts can produce star forming activity. The upper shaded region in Fig.~\ref{R_M} indicates the parameter space of massive protoclusters, defined by \cite{2012ApJ...758L..28B}, where there is almost no WGO distribution. Based on theoretical arguments,  \cite{2008Natur.451.1082K} establishs a critical value of $\Sigma_{\rm crit} = 1~\rm g~\rm cm^{-2}$ ($M(r) = \pi \Sigma_{\rm crit}r^{2}$, dash-dotted yellow line in Fig.~\ref{R_M}), but \cite{2012ApJ...754....5B} and \cite{2010A&A...517A..66L} obtained a smaller value of $\Sigma_{\rm crit} = 0.3~\rm g~\rm cm^{-2}$. \cite{2010ApJ...723L...7K} propose an empirical threshold $M(r) \geqslant 870\ M_{\odot}(r/\rm{pc})^{1.33}$ (dashed red line) as a minimum condition for massive star formation. 

%So mainly massive protostar luminosity, low-mass protostars contribute little.
Columns (9) and (10) in Table \ref{tab2} and \ref{tab3} mark whether robust or candidate MYSOs are greater than thresholds of $M(r) = 870\ M_{\odot}(r/\rm pc)^{1.33}$ (hereafter threshold one) and $\Sigma_{\rm crit} = 1~\rm g/\rm cm^{2}$ (hereafter threshold two). Robust and candidate MYSOs almost all both exceed the threshold one but show significant differences under the threshold two criteria. About 36.7\% of candidate MYSOs do not meet threshold two, but this proportion is only 22.5\% of robust MYSOs. However, this ratio was significantly increased among non MYSO candidates, and nearly 60\% of them did not exceed the threshold two. We infer that the reason for this difference in the ratios is that higher density gas may provide higher accretion just like robust MYSOs, and lower gas mass and density are also difficult to provide larger accretion rate, thus forming MYSOs. Column (11) in Table \ref{tab2} and \ref{tab3} is mark that sources whether are associated with Class II methanol maser that is one of the most reliable and sensitive tracers in the early stage of high-mass star formation region (\citealt{2006ApJ...638..241E}). Nearly 50\% of robust MYSOs are associated with Class II methanol maser, and that about 27\% of candidate MYSOs, which shows that the former as the MYSOs are more reliable than the latter.
%Difference in association rates between two type WGO MYSOs and Class II methanol maser 
%would further infer the reliability of these two catalogs.
%WGOs which listed in Table \ref{tab2} and \ref{tab3} are in deep blue and light yellow area.
%The green dashed line shows mass of WGOs whether greater or less than 1 $M_{\odot}$

%high mass 
\section{Conclusions}\label{CONCLUSIONS}
We screened out the WISE objects with green features in the whole Galactic Plane ($0^\circ<l<360^\circ$ and $-2^\circ<b<2^\circ$) and selected MYSOs sample with the help of SED fitting from mid-infrared to far-infrared and gravity thresholds. Our findings can be summarized as follows:
\begin{itemize}
\item
After cross-matching with the SIMBAD database to ensure non-YSO exclusions, we identified 2135 WGOs and divided them into three groups. The first group comprises 264 WGOs with an extended green structure, similar to the traditional {\it Spitzer} EGOs. 1366 WGOs without extended structures but showing compact green features are in the second group. The third group of 505 WGOs is neither extended nor visually green, but numerically W2(green)/W1(blue) is greater than 4.5. %The physical characteristics of Groups 2 and 3 suggest that they are at an earlier evolutionary stage than the first group. 
\item  
We find that $\sim 75$\% WGOs are distributed in $\mid l \mid <60$\degree, and in $\mid l \mid >60$\degree~the distribution of the WGOs is consistent with known star-forming regions. The distribution of WGOs is positively correlated with the density of molecular gas, except in the ultra-dense gas region of the galactic center. The WGOs have a Gaussian distribution along the galactic latitude ($\mid b \mid<2$\degree) but shift to negative Galactic latitude by $\sim 0.1$\degree. 
%and can trace the thin disk of the Milky Way 
%within 6\arcsecin its range
\item 
1260 WGOs have Hi-GAL source counterparts, accounting for 71\% of the total in Hi-GAL survey range. {\it Spitzer/IRAS} survey catalog shows a high association rate similar to that Hi-GAL catalog, over 54\% WGOs matched successfully. However, WISE \& 2MASS, RMS, ATLASGAL survey catalog for YSOs show very low association rate with WGOs, 3.2\%, 11.1\%, 20.8\% respectively. By cross-matching with these three catalogs, we obtained 2067, 1635, and 1176 YSOs newly identified and 348 new MYSOs from RMS MYSO sub-catalog.
\item
The WGOs in Group 1 are in the later evolutionary stage compared to the other two groups, as they have extended structures that is characteristic of outflow and a significantly higher rate of association with various masers, in addition physical parameters such as the luminosity of the parent structure and the gas accretion rate  are significantly higher.
%After the analysis of the WGOs in Group 2, we infer that they are younger than Group 3.
%are implied to have the longest time scale
%as it is associated not only with the highest rate of various masers, but also with the largest number of masers at the same time.
%The association rate between WGOs and various masers may provide evidence that they are in different evolutionary sequences.
\item 
From SED fitting, we obtained the stellar mass $M_{\star}$ and accretion rate $\dot M_{\rm acc}$ of 1151 WGOs. The luminosity of WGOs in goups one is significantly larger than the other two. 231 WGOs with stellar mass $M_{\star}+\dot M_{\rm acc} \times 10^{4}\ {\rm yr}>8 \ M_{\odot}$ are selected as robust MYSOs. 172 WGOs with accretion rate $\dot M_{\rm acc}>10^{-4}\ M_{\odot}\rm{yr}^{-1}$ are selected as candidate MYSOs. We find that accretion rate and the density of the parent structure of WGOs are positively correlated.
\end{itemize}
\renewcommand\arraystretch{7}
\setlength{\tabcolsep}{4pt}
\setlength{\LTleft}{0pt}

\startlongtable
\begin{deluxetable*}{ccccccccccc}
\tablenum{2}
\tablecaption{Robust MYSOs \label{tab2}}
\tablewidth{0pt}
\tabletypesize{\scriptsize}
\tablehead{
\colhead{Name}  & \colhead{Group} & 
\colhead{$l$} & \colhead{$b$} & \colhead{Model set} & \colhead{Stellar mass} &
\colhead{Accretion rate} & \colhead{Distance} & \colhead{$\geqslant 870\ M_{\odot} R^{1.33}$} & \colhead{$\geqslant 1~\rm g/cm^{2}$} & \colhead{Class II}\\
{}& {} & {} & 
{} &{}& {($M_{\star}$)} &{($\dot M_{\rm acc}$)} & {}  & {} & {} & {methanol maser}\\
\colhead{}& {} & \dcolhead{^{\circ}} & 
\dcolhead{^{\circ}} &\colhead{}& \dcolhead{M_{\odot}} &\dcolhead{\times 10^{-4} M_{\odot}\rm yr^{-1}} & \colhead{kpc}  & \colhead{} & \colhead{} & \colhead{}} 
\decimalcolnumbers
\startdata
G000.091-00.663 & I & 0.0912 & -0.6628 & s-ubsmi & 3.17(0.48) & 6.073 & 8.21 & $\surd$ & $\surd$ & $\surd$ \\
G000.484-00.700 & I & 0.4837 & -0.7003 & s-u-smi & 15.72(4.83) & 3.684 & 7.32 & $\surd$ & $\surd$ & {} \\
G002.529+00.199 & I & 2.529 & 0.199 & s-u-hmi & 3.71(0.04) & 31.305 & 12.91 & $\surd$ & $\surd$ & {} \\
G004.827+00.231 & I & 4.827 & 0.2306 & s-pbhmi & 2.23(0.0) & 55.936 & 3.62 & $\surd$ & $\surd$ & {} \\
G006.797-00.258 & I & 6.7968 & -0.2581 & s-ubsmi & 2.56(0.21) & 116.692 & 3.82 & $\surd$ & $\surd$ & $\surd$ \\
G009.779-00.167 & I & 9.7785 & -0.1672 & s-u-smi & 4.28(3.82) & 17.838 & 11.92 & $\surd$ & {} & {} \\
G010.628-00.337 & I & 10.6281 & -0.3369 & s-ubhmi & 5.43(4.24) & 68.455 & 17.23 & $\surd$ & {} & {} \\
G019.008-00.029 & I & 19.0084 & -0.0295 & s-ubhmi & 2.8(0.58) & 315.365 & 11.67 & $\surd$ & $\surd$ & $\surd$ \\
G019.884-00.535 & I & 19.8837 & -0.5351 & s-ubsmi & 2.43(0.0) & 45.547 & 3.25 & $\surd$ & $\surd$ & $\surd$ \\
G019.977-00.214 & I & 19.9773 & -0.2141 & s-u-smi & 2.24(0.81) & 8.53 & 12.35 & $\surd$ & $\surd$ & {} \\
G027.795-00.277 & I & 27.7951 & -0.2771 & s-u-smi & 2.11(0.69) & 7.524 & 2.94 & $\surd$ & $\surd$ & {} \\
G027.968-00.475 & I & 27.9676 & -0.4746 & s-u-smi & 1.99(0.53) & 30.533 & 11.82 & $\surd$ & $\surd$ & {} \\
G030.652-00.204 & I & 30.6516 & -0.2042 & s-u-hmi & 3.67(0.0) & 35.535 & 9.3 & $\surd$ & $\surd$ & $\surd$ \\
G030.788+00.203 & I & 30.7876 & 0.2032 & s-u-smi & 4.53(4.04) & 4.568 & 9.71 & $\surd$ & $\surd$ & $\surd$ \\
G031.580+00.076 & I & 31.5801 & 0.0757 & s-pbhmi & 2.65(0.0) & 49.676 & 4.9 & $\surd$ & $\surd$ & $\surd$ \\
G039.388-00.141 & I & 39.3877 & -0.1414 & s-u-smi & 2.24(0.0) & 15.235 & 4.0 & $\surd$ & $\surd$ & $\surd$ \\
G039.495-00.993 & I & 39.4948 & -0.993 & s-u-hmi & 3.57(4.81) & 8.559 & 3.28 & $\surd$ & {} & {} \\
G040.278-00.269 & I & 40.2781 & -0.269 & s-pbhmi & 2.92(2.42) & 11.172 & 8.17 & $\surd$ & {} & $\surd$ \\
G045.465+00.043 & I & 45.4652 & 0.0431 & s-u-smi & 3.31(0.66) & 12.322 & 7.08 & $\surd$ & $\surd$ & {} \\
G045.468+00.048 & I & 45.4678 & 0.0484 & s-u-smi & 3.3(0.72) & 12.458 & 7.08 & $\surd$ & $\surd$ & {} \\
G058.468+00.437 & I & 58.4675 & 0.4366 & s-pbhmi & 9.72(18.36) & 0.287 & 4.36 & $\surd$ & $\surd$ & {} \\
G059.783+00.065 & I & 59.7827 & 0.0649 & s-u-hmi & 1.96(0.0) & 10.637 & 2.1 & $\surd$ & $\surd$ & $\surd$ \\
G077.462+01.760 & I & 77.462 & 1.7596 & s-u-smi & 7.55(1.68) & 12.341 & 3.35 & $\surd$ & $\surd$ & {} \\
G311.513-00.454 & I & 311.5131 & -0.4536 & s-u-smi & 3.2(2.38) & 9.513 & 3.68 & $\surd$ & $\surd$ & {} \\
G313.705-00.190 & I & 313.7051 & -0.1899 & s-u-hmi & 3.43(1.88) & 70.496 & 8.69 & $\surd$ & $\surd$ & $\surd$ \\
G313.767-00.862 & I & 313.7668 & -0.8625 & spu-hmi & 3.56(0.0) & 267.344 & 8.08 & $\surd$ & $\surd$ & $\surd$ \\
G316.640-00.087 & I & 316.6403 & -0.0867 & s-u-smi & 4.28(4.24) & 37.478 & 10.86 & $\surd$ & $\surd$ & $\surd$ \\
G316.959+00.302 & I & 316.9588 & 0.3022 & s-u-smi & 3.93(1.19) & 104.254 & 9.61 & $\surd$ & $\surd$ & {} \\
G317.466-00.403 & I & 317.4658 & -0.4026 & s-u-hmi & 2.87(0.0) & 30.673 & 9.58 & $\surd$ & $\surd$ & $\surd$ \\
G318.948-00.196 & I & 318.9476 & -0.196 & s-ubsmi & 12.55(2.68) & 22.042 & 10.38 & $\surd$ & $\surd$ & $\surd$ \\
G320.233-00.284 & I & 320.2328 & -0.2843 & s-u-hmi & 2.79(0.0) & 208.682 & 8.6 & $\surd$ & $\surd$ & $\surd$ \\
G320.892-00.411 & I & 320.8923 & -0.411 & s-u-smi & 4.02(0.0) & 139.108 & 10.2 & $\surd$ & $\surd$ & {} \\
G324.716+00.341 & I & 324.7161 & 0.3414 & s-u-smi & 3.36(0.89) & 7.448 & 10.46 & $\surd$ & $\surd$ & $\surd$ \\
G326.544+00.169 & I & 326.5442 & 0.1693 & spu-hmi & 3.09(1.26) & 32.043 & 4.31 & $\surd$ & $\surd$ & {} \\
G326.780-00.241 & I & 326.78 & -0.241 & s-pbhmi & 3.37(1.63) & 20.657 & 3.83 & $\surd$ & $\surd$ & {} \\
G328.824-00.080 & I & 328.8237 & -0.0798 & s-u-smi & 3.82(0.0) & 11.54 & 12.08 & $\surd$ & {} & {} \\
G330.070+01.063 & I & 330.07 & 1.0634 & s-ubsmi & 6.48(7.74) & 65.722 & 11.78 & $\surd$ & $\surd$ & $\surd$ \\
G332.352-00.117 & I & 332.3522 & -0.1172 & s-u-hmi & 1.98(0.0) & 7.846 & 3.15 & $\surd$ & $\surd$ & $\surd$ \\
G332.364+00.606 & I & 332.364 & 0.6057 & s-u-smi & 3.44(0.0) & 27.394 & 12.0 & $\surd$ & $\surd$ & $\surd$ \\
G335.060-00.427 & I & 335.0595 & -0.4274 & s-pbsmi & 2.22(1.42) & 23.016 & 2.72 & $\surd$ & $\surd$ & $\surd$ \\
G337.299-00.874 & I & 337.2993 & -0.8741 & s-pbsmi & 2.75(0.62) & 28.342 & 10.25 & $\surd$ & $\surd$ & $\surd$ \\
G338.280+00.542 & I & 338.28 & 0.5423 & s-ubhmi & 2.53(0.0) & 5.829 & 3.93 & $\surd$ & $\surd$ & $\surd$ \\
G339.584-00.127 & I & 339.5838 & -0.1273 & sp--smi & 3.91(2.77) & 5.785 & 12.91 & $\surd$ & {} & $\surd$ \\
G339.927-00.083 & I & 339.9269 & -0.0829 & s-ubhmi & 3.04(0.17) & 21.137 & 3.76 & $\surd$ & $\surd$ & {} \\
G342.705+00.127 & I & 342.7053 & 0.1265 & s-u-hmi & 5.54(0.0) & 104.625 & 12.69 & $\surd$ & {} & {} \\
G352.315-00.442 & I & 352.3155 & -0.4421 & s-pbhmi & 2.41(0.0) & 5.698 & 2.0 & $\surd$ & $\surd$ & {} \\
G000.084-00.642 & II & 0.084 & -0.6417 & s-pbhmi & 2.34(0.06) & 35.679 & 8.46 & $\surd$ & $\surd$ & {} \\
G000.547-00.851 & II & 0.5468 & -0.8509 & s-u-hmi & 4.17(0.0) & 216.786 & 7.43 & $\surd$ & $\surd$ & $\surd$ \\
G001.147-00.125 & II & 1.1468 & -0.1247 & s-pbsmi & 5.59(1.65) & 58.091 & 11.23 & $\surd$ & $\surd$ & $\surd$ \\
G007.470+00.057 & II & 7.4701 & 0.0575 & s-ubhmi & 5.17(3.67) & 30.284 & 14.12 & $\surd$ & $\surd$ & {} \\
G007.601-00.139 & II & 7.6007 & -0.139 & spu-smi & 3.6(0.63) & 135.516 & 8.94 & $\surd$ & $\surd$ & $\surd$ \\
G007.635-00.192 & II & 7.6347 & -0.1923 & s-pbhmi & 1.78(0.0) & 7.134 & 8.93 & $\surd$ & $\surd$ & {} \\
G008.412-00.346 & II & 8.4118 & -0.3455 & s-ubhmi & 3.42(1.06) & 6.602 & 12.02 & $\surd$ & {} & {} \\
G008.722-00.404 & II & 8.7219 & -0.4036 & s-pbsmi & 3.49(4.76) & 5.982 & 12.05 & $\surd$ & $\surd$ & {} \\
G010.621-00.443 & II & 10.6214 & -0.4429 & s-ubsmi & 3.93(0.79) & 39.08 & 16.72 & $\surd$ & {} & {} \\
G010.622-00.443 & II & 10.6221 & -0.4434 & s-ubsmi & 3.93(0.79) & 39.08 & 16.72 & $\surd$ & {} & {} \\
G010.724-00.335 & II & 10.724 & -0.3347 & s-u-hmi & 4.59(2.21) & 97.942 & 16.69 & $\surd$ & {} & $\surd$ \\
G011.072-00.386 & II & 11.0717 & -0.3857 & s-pbsmi & 3.25(2.98) & 51.76 & 16.55 & $\surd$ & {} & {} \\
G011.109-00.114 & II & 11.1089 & -0.1143 & s-pbsmi & 3.26(0.6) & 279.858 & 13.07 & $\surd$ & $\surd$ & $\surd$ \\
G015.665-00.499 & II & 15.6653 & -0.4986 & s-u-smi & 3.63(0.0) & 7.774 & 14.31 & $\surd$ & $\surd$ & $\surd$ \\
G017.736-00.241 & II & 17.7356 & -0.2409 & s-ubhmi & 2.85(0.04) & 31.171 & 10.18 & $\surd$ & {} & {} \\
G018.262-00.244 & II & 18.2618 & -0.2437 & s-u-hmi & 2.2(1.15) & 9.581 & 4.51 & $\surd$ & $\surd$ & $\surd$ \\
G018.628-00.070 & II & 18.6275 & -0.0702 & s-u-smi & 2.01(1.96) & 6.678 & 12.42 & $\surd$ & {} & {} \\
G019.611-00.120 & II & 19.6109 & -0.1202 & s-ubhmi & 2.89(0.95) & 22.682 & 11.77 & $\surd$ & $\surd$ & {} \\
G019.828-00.330 & II & 19.8284 & -0.33 & spu-smi & 5.02(6.92) & 24.731 & 12.39 & $\surd$ & {} & {} \\
G022.356+00.066 & II & 22.3558 & 0.0663 & spu-smi & 2.88(2.8) & 10.917 & 4.85 & $\surd$ & $\surd$ & $\surd$ \\
G023.009-00.411 & II & 23.0095 & -0.4108 & s-pbsmi & 2.97(2.03) & 152.708 & 4.6 & $\surd$ & $\surd$ & $\surd$ \\
G023.198-00.001 & II & 23.1977 & 0.0006 & sp--smi & 6.35(8.15) & 10.881 & 10.81 & $\surd$ & $\surd$ & {} \\
G023.965-00.110 & II & 23.9648 & -0.1103 & s-pbsmi & 2.71(0.0) & 11.232 & 4.39 & $\surd$ & $\surd$ & $\surd$ \\
G023.994-00.097 & II & 23.9944 & -0.0973 & s-u-smi & 6.91(8.765) & 15.961 & 11.0 & $\surd$ & $\surd$ & {} \\
G024.148-00.009 & II & 24.1483 & -0.0092 & s-ubsmi & 7.32(1.13) & 17.691 & 10.32 & $\surd$ & $\surd$ & $\surd$ \\
G024.427+00.122 & II & 24.4269 & 0.1222 & s-pbhmi & 3.05(0.0) & 10.19 & 9.24 & $\surd$ & $\surd$ & {} \\
G024.427+00.123 & II & 24.4269 & 0.1226 & s-pbhmi & 3.05(0.0) & 10.19 & 9.24 & $\surd$ & $\surd$ & {} \\
G024.541+00.312 & II & 24.5407 & 0.3121 & s-pbhmi & 2.89(1.81) & 6.458 & 5.79 & $\surd$ & {} & $\surd$ \\
G024.626-00.101 & II & 24.6259 & -0.1009 & s-ubhmi & 3.3(2.08) & 7.604 & 9.18 & $\surd$ & $\surd$ & {} \\
G024.731+00.154 & II & 24.7306 & 0.1541 & s-u-hmi & 2.78(0.79) & 42.658 & 5.81 & $\surd$ & $\surd$ & {} \\
G025.710+00.044 & II & 25.7095 & 0.0443 & s-pbsmi & 4.32(3.12) & 26.534 & 10.2 & $\surd$ & $\surd$ & $\surd$ \\
G027.016+00.200 & II & 27.0162 & 0.2001 & s-u-hmi & 2.94(0.15) & 8.217 & 9.74 & $\surd$ & $\surd$ & {} \\
G027.248+00.108 & II & 27.2476 & 0.1081 & s-ubsmi & 1.55(0.23) & 7.171 & 9.87 & $\surd$ & {} & {} \\
G028.597-00.021 & II & 28.5967 & -0.0209 & sp--smi & 6.23(1.95) & 3.293 & 9.14 & $\surd$ & {} & {} \\
G028.652+00.027 & II & 28.6519 & 0.0268 & s-pbsmi & 9.73(4.39) & 34.893 & 8.97 & $\surd$ & $\surd$ & {} \\
G028.700+00.406 & II & 28.6998 & 0.4063 & s-ubsmi & 4.49(4.87) & 5.17 & 9.55 & $\surd$ & {} & $\surd$ \\
G028.842+00.494 & II & 28.8424 & 0.4936 & s-pbhmi & 3.01(0.0) & 7.697 & 9.89 & $\surd$ & $\surd$ & $\surd$ \\
G030.348+00.392 & II & 30.3478 & 0.392 & s-pbhmi & 2.53(0.0) & 54.637 & 9.24 & $\surd$ & $\surd$ & {} \\
G030.603+00.176 & II & 30.6035 & 0.1758 & s-u-hmi & 3.18(0.0) & 8.113 & 8.45 & $\surd$ & $\surd$ & {} \\
G030.770-00.804 & II & 30.7696 & -0.8044 & s-ubsmi & 2.27(0.75) & 43.502 & 9.87 & $\surd$ & $\surd$ & $\surd$ \\
G030.770-00.805 & II & 30.7697 & -0.8048 & s-ubsmi & 2.27(0.75) & 43.502 & 9.87 & $\surd$ & $\surd$ & $\surd$ \\
G030.920+00.088 & II & 30.9203 & 0.0878 & s-ubhmi & 2.59(0.78) & 11.419 & 8.96 & $\surd$ & $\surd$ & {} \\
G030.972-00.142 & II & 30.9723 & -0.1417 & s-ubhmi & 2.63(0.44) & 7.386 & 4.44 & $\surd$ & $\surd$ & $\surd$ \\
G031.900+00.341 & II & 31.9003 & 0.3407 & s-u-hmi & 1.66(0.7) & 13.185 & 14.06 & $\surd$ & {} & {} \\
G032.118+00.090 & II & 32.1177 & 0.0903 & s-u-hmi & 3.34(0.0) & 27.84 & 8.37 & $\surd$ & $\surd$ & $\surd$ \\
G032.992+00.034 & II & 32.9917 & 0.034 & s-pbsmi & 3.56(0.0) & 29.694 & 9.26 & $\surd$ & $\surd$ & $\surd$ \\
G033.229-00.018 & II & 33.2292 & -0.0184 & s-u-smi & 1.99(0.0) & 26.258 & 8.02 & $\surd$ & $\surd$ & $\surd$ \\
G033.393+00.010 & II & 33.3927 & 0.0096 & s-pbhmi & 3.02(0.13) & 7.732 & 7.83 & $\surd$ & $\surd$ & $\surd$ \\
G035.148+00.809 & II & 35.1484 & 0.8091 & s-pbsmi & 2.01(0.0) & 21.779 & 9.29 & $\surd$ & $\surd$ & $\surd$ \\
G035.417-00.285 & II & 35.4167 & -0.2846 & s-pbhmi & 2.1(0.0) & 97.352 & 10.46 & $\surd$ & $\surd$ & {} \\
G037.769-00.393 & II & 37.769 & -0.3925 & s-ubhmi & 2.29(0.71) & 7.821 & 12.33 & $\surd$ & {} & {} \\
G037.874-00.399 & II & 37.8736 & -0.3991 & s-u-hmi & 8.83(5.21) & 34.673 & 9.45 & $\surd$ & $\surd$ & {} \\
G038.120-00.228 & II & 38.1195 & -0.2279 & s-pbhmi & 2.74(0.0) & 28.125 & 5.24 & $\surd$ & $\surd$ & $\surd$ \\
G038.933-00.361 & II & 38.9331 & -0.3607 & s-u-hmi & 2.85(1.41) & 8.434 & 10.48 & $\surd$ & $\surd$ & {} \\
G039.100+00.491 & II & 39.0999 & 0.4909 & s-ubhmi & 2.73(0.0) & 28.917 & 11.48 & $\surd$ & $\surd$ & $\surd$ \\
G041.307-00.169 & II & 41.3067 & -0.1693 & s-pbhmi & 1.98(0.0) & 21.493 & 8.96 & $\surd$ & {} & $\surd$ \\
G043.037-00.451 & II & 43.0374 & -0.4512 & s-ubsmi & 4.1(3.88) & 71.495 & 8.5 & $\surd$ & $\surd$ & $\surd$ \\
G043.074-00.077 & II & 43.0742 & -0.0775 & s-u-hmi & 1.69(0.0) & 14.642 & 11.32 & $\surd$ & $\surd$ & $\surd$ \\
G043.177-00.518 & II & 43.1771 & -0.5185 & s-pbsmi & 3.61(0.3) & 34.346 & 8.46 & $\surd$ & $\surd$ & $\surd$ \\
G048.364+00.247 & II & 48.3635 & 0.2469 & s-ubsmi & 1.59(0.13) & 12.945 & 10.66 & $\surd$ & {} & {} \\
G048.856+00.235 & II & 48.8556 & 0.2351 & s-ubhmi & 2.98(0.65) & 10.104 & 9.9 & $\surd$ & {} & {} \\
G050.033+00.581 & II & 50.0335 & 0.5811 & s-u-smi & 3.12(0.25) & 26.955 & 10.85 & $\surd$ & $\surd$ & $\surd$ \\
G054.110-00.081 & II & 54.1097 & -0.0812 & s-ubsmi & 2.61(1.75) & 17.184 & 4.3 & $\surd$ & $\surd$ & {} \\
G071.312+00.828 & II & 71.3124 & 0.8278 & spu-hmi & 3.95(3.8) & 62.095 & 4.08 & $\surd$ & $\surd$ & {} \\
G111.859+00.801 & II & 111.8595 & 0.8013 & s-u-hmi & 2.34(0.0) & 16.961 & 3.2 & $\surd$ & $\surd$ & {} \\
G286.727-00.194 & II & 286.7271 & -0.1942 & s-pbsmi & 2.45(0.0) & 12.148 & 6.32 & $\surd$ & $\surd$ & {} \\
G297.140-01.341 & II & 297.14 & -1.3406 & s-u-smi & 2.97(3.05) & 12.469 & 9.07 & $\surd$ & {} & {} \\
G297.403-00.652 & II & 297.4026 & -0.6515 & s-pbsmi & 2.53(0.05) & 10.158 & 10.01 & $\surd$ & {} & {} \\
G298.723-00.086 & II & 298.7231 & -0.0861 & s-pbsmi & 2.45(0.0) & 9.799 & 4.01 & $\surd$ & $\surd$ & $\surd$ \\
G311.567+00.319 & II & 311.5671 & 0.3186 & sp--hmi & 2.95(1.8) & 20.972 & 7.65 & $\surd$ & $\surd$ & {} \\
G311.627+00.290 & II & 311.6269 & 0.29 & s-ubsmi & 3.77(1.29) & 5.461 & 6.88 & $\surd$ & $\surd$ & $\surd$ \\
G311.629+00.266 & II & 311.6286 & 0.2655 & s-ubhmi & 3.26(3.16) & 9.879 & 4.64 & $\surd$ & $\surd$ & $\surd$ \\
G311.947+00.142 & II & 311.9467 & 0.1417 & s-u-hmi & 2.79(0.0) & 7.407 & 8.29 & $\surd$ & $\surd$ & $\surd$ \\
G312.412-01.050 & II & 312.4121 & -1.0496 & s-u-hmi & 2.35(1.24) & 6.843 & 3.14 & $\surd$ & $\surd$ & {} \\
G312.590-00.503 & II & 312.5899 & -0.5027 & s-u-hmi & 2.64(0.0) & 23.437 & 6.79 & $\surd$ & {} & {} \\
G313.469+00.190 & II & 313.4692 & 0.1904 & spu-smi & 5.02(5.75) & 102.515 & 11.01 & $\surd$ & $\surd$ & $\surd$ \\
G314.239+00.365 & II & 314.2387 & 0.3653 & s-ubsmi & 2.76(1.83) & 9.824 & 6.7 & $\surd$ & {} & {} \\
G317.028+00.361 & II & 317.0283 & 0.3613 & spu-smi & 2.77(0.23) & 12.177 & 9.03 & $\surd$ & {} & $\surd$ \\
G319.178-00.409 & II & 319.1783 & -0.4085 & spu-hmi & 4.92(5.33) & 60.745 & 10.93 & $\surd$ & {} & {} \\
G326.268-00.486 & II & 326.2685 & -0.4858 & s-pbhmi & 4.06(3.1) & 101.904 & 10.31 & $\surd$ & $\surd$ & {} \\
G326.279+00.561 & II & 326.2787 & 0.5612 & s-u-smi & 1.66(0.0) & 6.539 & 11.19 & $\surd$ & {} & {} \\
G326.322-00.395 & II & 326.3222 & -0.3953 & s-pbhmi & 2.53(0.0) & 58.885 & 9.69 & $\surd$ & $\surd$ & $\surd$ \\
G326.795+00.382 & II & 326.7951 & 0.3822 & s-u-hmi & 3.69(0.0) & 40.486 & 12.53 & $\surd$ & $\surd$ & {} \\
G329.466+00.516 & II & 329.4664 & 0.5157 & sp--hmi & 3.16(0.46) & 7.236 & 10.42 & $\surd$ & $\surd$ & {} \\
G329.717+00.804 & II & 329.7174 & 0.8038 & s-pbhmi & 1.39(0.09) & 12.89 & 11.32 & $\surd$ & $\surd$ & {} \\
G329.719+01.164 & II & 329.7189 & 1.1638 & s-ubsmi & 2.07(0.0) & 15.379 & 4.45 & $\surd$ & $\surd$ & $\surd$ \\
G330.026+01.043 & II & 330.0258 & 1.0428 & s-ubsmi & 1.69(0.0) & 9.026 & 11.49 & $\surd$ & $\surd$ & {} \\
G330.035+01.045 & II & 330.035 & 1.0446 & s-ubhmi & 1.67(0.52) & 9.785 & 11.47 & $\surd$ & $\surd$ & {} \\
G331.119-00.118 & II & 331.1191 & -0.1181 & s-ubsmi & 2.37(0.54) & 14.238 & 9.79 & $\surd$ & {} & $\surd$ \\
G331.342-00.346 & II & 331.3415 & -0.3462 & s-ubsmi & 11.51(3.73) & 12.684 & 3.88 & $\surd$ & $\surd$ & $\surd$ \\
G331.491-00.116 & II & 331.491 & -0.1157 & s-pbhmi & 2.84(1.39) & 54.36 & 4.82 & $\surd$ & $\surd$ & {} \\
G332.583+00.147 & II & 332.5827 & 0.1473 & s-pbsmi & 2.76(1.76) & 6.042 & 11.89 & $\surd$ & $\surd$ & $\surd$ \\
G332.942-00.686 & II & 332.9415 & -0.6863 & s-ubsmi & 4.47(4.45) & 5.961 & 3.8 & $\surd$ & $\surd$ & {} \\
G333.184-00.091 & II & 333.1838 & -0.0907 & s-ubsmi & 2.08(0.0) & 15.636 & 4.85 & $\surd$ & $\surd$ & $\surd$ \\
G336.410-00.255 & II & 336.4102 & -0.2554 & s-u-hmi & 5.99(5.49) & 79.671 & 10.34 & $\surd$ & $\surd$ & $\surd$ \\
G336.432-00.262 & II & 336.4325 & -0.2616 & s-pbhmi & 2.85(0.68) & 137.634 & 10.22 & $\surd$ & $\surd$ & $\surd$ \\
G336.871+00.295 & II & 336.8706 & 0.2954 & s-u-hmi & 1.57(0.0) & 17.249 & 11.21 & $\surd$ & $\surd$ & {} \\
G336.941-00.156 & II & 336.9405 & -0.1562 & s-ubsmi & 3.23(0.64) & 107.585 & 10.96 & $\surd$ & $\surd$ & $\surd$ \\
G336.958-00.977 & II & 336.9578 & -0.9771 & s-ubhmi & 3.56(0.68) & 31.592 & 12.22 & $\surd$ & $\surd$ & $\surd$ \\
G337.686+00.138 & II & 337.6859 & 0.1382 & s-pbhmi & 2.58(1.18) & 41.125 & 11.01 & $\surd$ & $\surd$ & $\surd$ \\
G338.140+00.178 & II & 338.1396 & 0.1783 & s-ubsmi & 2.51(1.3) & 43.855 & 13.02 & $\surd$ & $\surd$ & $\surd$ \\
G338.392-00.403 & II & 338.3916 & -0.4029 & spu-smi & 4.15(2.25) & 87.926 & 12.72 & $\surd$ & $\surd$ & $\surd$ \\
G338.472+00.289 & II & 338.4718 & 0.2887 & s-pbsmi & 2.2(1.29) & 8.928 & 2.37 & $\surd$ & $\surd$ & $\surd$ \\
G339.948-00.539 & II & 339.9484 & -0.5389 & s-ubsmi & 2.84(0.0) & 42.258 & 5.32 & $\surd$ & $\surd$ & $\surd$ \\
G341.215-00.236 & II & 341.2152 & -0.2358 & s-u-hmi & 5.58(2.35) & 14.925 & 3.45 & $\surd$ & $\surd$ & {} \\
G341.218-00.213 & II & 341.2178 & -0.2127 & spu-smi & 2.43(0.89) & 47.365 & 3.37 & $\surd$ & $\surd$ & $\surd$ \\
G342.905-00.120 & II & 342.9048 & -0.1202 & s-pbhmi & 2.41(0.0) & 6.136 & 5.25 & $\surd$ & $\surd$ & {} \\
G343.444-00.203 & II & 343.444 & -0.2032 & s-ubsmi & 2.31(1.35) & 32.268 & 16.36 & $\surd$ & {} & {} \\
G344.065+00.114 & II & 344.0655 & 0.1135 & s-u-hmi & 2.71(0.88) & 5.7 & 11.47 & $\surd$ & {} & {} \\
G345.985-00.020 & II & 345.9847 & -0.0205 & s-u-smi & 3.09(0.04) & 17.216 & 10.81 & $\surd$ & $\surd$ & $\surd$ \\
G346.369-00.647 & II & 346.3686 & -0.6471 & s-u-hmi & 1.93(0.0) & 11.93 & 17.06 & $\surd$ & {} & {} \\
G347.817+00.018 & II & 347.8168 & 0.0177 & s-u-hmi & 1.57(0.0) & 17.244 & 13.06 & $\surd$ & $\surd$ & $\surd$ \\
G350.105+00.080 & II & 350.1047 & 0.0804 & s-pbhmi & 5.06(5.25) & 9.29 & 5.6 & $\surd$ & $\surd$ & {} \\
G352.161+00.199 & II & 352.1612 & 0.1992 & spu-hmi & 2.35(0.4) & 24.291 & 11.55 & $\surd$ & $\surd$ & {} \\
G357.553-00.548 & II & 357.553 & -0.5477 & s-pbhmi & 4.27(0.0) & 103.827 & 17.11 & $\surd$ & {} & {} \\
G357.609-00.618 & II & 357.6093 & -0.618 & s-u-hmi & 4.36(1.59) & 78.327 & 18.9 & $\surd$ & {} & {} \\
G357.966-00.163 & II & 357.9664 & -0.1634 & sp--smi & 5.03(0.67) & 9.241 & 13.95 & $\surd$ & $\surd$ & $\surd$ \\
G359.436-00.102 & II & 359.4363 & -0.1018 & s-u-hmi & 3.77(0.0) & 146.105 & 8.62 & $\surd$ & $\surd$ & $\surd$ \\
G007.333-00.016 & III & 7.3335 & -0.0163 & spu-smi & 2.95(1.69) & 20.539 & 13.21 & $\surd$ & $\surd$ & {} \\
G010.575-00.577 & III & 10.5749 & -0.5771 & s-ubsmi & 5.67(9.27) & 281.369 & 16.4 & $\surd$ & $\surd$ & {} \\
G010.824-00.103 & III & 10.8238 & -0.1031 & s-u-smi & 3.17(0.67) & 37.947 & 11.0 & $\surd$ & {} & $\surd$ \\
G011.116+00.051 & III & 11.1156 & 0.0514 & s-u-smi & 5.34(6.54) & 30.111 & 13.48 & $\surd$ & $\surd$ & {} \\
G011.992-00.272 & III & 11.9917 & -0.2722 & s-pbhmi & 3.18(0.0) & 58.651 & 11.44 & $\surd$ & $\surd$ & $\surd$ \\
G012.025-00.032 & III & 12.0252 & -0.0316 & s-u-smi & 3.4(2.12) & 139.828 & 9.4 & $\surd$ & $\surd$ & $\surd$ \\
G012.158-00.134 & III & 12.1583 & -0.1343 & s-ubhmi & 3.56(0.0) & 247.357 & 16.2 & $\surd$ & {} & {} \\
G012.199-00.034 & III & 12.1986 & -0.0342 & s-pbhmi & 4.72(0.0) & 95.534 & 11.81 & $\surd$ & $\surd$ & $\surd$ \\
G012.525+00.016 & III & 12.5255 & 0.0158 & s-ubhmi & 3.56(1.51) & 96.208 & 12.84 & $\surd$ & $\surd$ & $\surd$ \\
G013.184-00.107 & III & 13.1836 & -0.1068 & spu-smi & 2.63(3.9) & 12.041 & 4.45 & $\surd$ & $\surd$ & {} \\
G013.210-00.143 & III & 13.2102 & -0.1434 & s-u-smi & 2.72(1.23) & 21.64 & 4.46 & $\surd$ & $\surd$ & {} \\
G013.713-00.084 & III & 13.7127 & -0.0837 & s-u-hmi & 1.89(0.3) & 16.938 & 4.05 & $\surd$ & $\surd$ & $\surd$ \\
G014.949-00.072 & III & 14.9493 & -0.0722 & s-pbsmi & 2.68(0.0) & 13.509 & 11.15 & $\surd$ & $\surd$ & {} \\
G017.964+00.080 & III & 17.9643 & 0.08 & s-u-smi & 13.89(6.63) & 56.429 & 13.84 & $\surd$ & $\surd$ & {} \\
G018.296+00.429 & III & 18.2964 & 0.4294 & s-ubsmi & 2.18(0.0) & 13.326 & 15.68 & {} & {} & {} \\
G020.239+00.065 & III & 20.2387 & 0.0649 & s-pbhmi & 2.1(0.0) & 12.094 & 4.43 & $\surd$ & $\surd$ & $\surd$ \\
G023.818+00.384 & III & 23.8176 & 0.3836 & spu-hmi & 4.63(5.1) & 32.933 & 10.76 & $\surd$ & {} & {} \\
G024.442-00.228 & III & 24.4424 & -0.2282 & s-u-smi & 2.06(0.47) & 26.987 & 3.69 & $\surd$ & $\surd$ & {} \\
G025.398+00.562 & III & 25.3979 & 0.5618 & s-pbsmi & 3.6(0.0) & 44.832 & 13.75 & $\surd$ & $\surd$ & {} \\
G025.613-00.137 & III & 25.613 & -0.1366 & s-u-hmi & 3.73(2.1) & 34.695 & 9.87 & $\surd$ & $\surd$ & {} \\
G026.496+00.711 & III & 26.4956 & 0.7105 & s-u-smi & 4.21(1.0) & 90.278 & 11.85 & $\surd$ & $\surd$ & {} \\
G029.319-00.162 & III & 29.3195 & -0.1617 & s-ubhmi & 3.12(0.33) & 36.091 & 11.59 & $\surd$ & $\surd$ & $\surd$ \\
G030.463+00.032 & III & 30.4633 & 0.0324 & s-pbsmi & 3.81(2.91) & 31.274 & 8.42 & $\surd$ & $\surd$ & {} \\
G030.588-00.043 & III & 30.5883 & -0.0427 & s-u-smi & 5.23(0.0) & 161.709 & 11.71 & $\surd$ & $\surd$ & $\surd$ \\
G030.866+00.114 & III & 30.866 & 0.1144 & s-u-smi & 5.75(4.27) & 66.956 & 11.9 & $\surd$ & $\surd$ & {} \\
G030.958+00.087 & III & 30.958 & 0.0866 & s-u-hmi & 4.24(2.08) & 502.697 & 11.76 & $\surd$ & $\surd$ & $\surd$ \\
G030.972+00.562 & III & 30.9723 & 0.5622 & s-pbsmi & 3.02(0.29) & 158.286 & 12.72 & $\surd$ & {} & $\surd$ \\
G030.980+00.216 & III & 30.9797 & 0.2156 & s-ubhmi & 4.42(5.81) & 7.426 & 8.07 & $\surd$ & $\surd$ & $\surd$ \\
G034.756+00.025 & III & 34.7563 & 0.0245 & s-u-hmi & 3.62(3.93) & 7.669 & 4.43 & $\surd$ & $\surd$ & $\surd$ \\
G036.878-00.473 & III & 36.8777 & -0.4729 & s-u-hmi & 2.39(0.54) & 16.941 & 3.66 & $\surd$ & $\surd$ & {} \\
G040.282-00.220 & III & 40.2816 & -0.2196 & s-pbhmi & 2.71(0.0) & 121.499 & 4.71 & $\surd$ & $\surd$ & $\surd$ \\
G041.132+00.129 & III & 41.132 & 0.1291 & spu-smi & 2.98(1.98) & 31.889 & 8.76 & $\surd$ & {} & {} \\
G042.098+00.352 & III & 42.0982 & 0.3517 & spu-hmi & 3.76(2.19) & 156.175 & 11.04 & $\surd$ & $\surd$ & {} \\
G043.306-00.211 & III & 43.3065 & -0.2108 & s-u-hmi & 2.29(0.0) & 20.407 & 3.84 & $\surd$ & $\surd$ & {} \\
G049.043-01.079 & III & 49.043 & -1.0786 & s-pbhmi & 5.13(5.95) & 26.783 & 8.2 & $\surd$ & $\surd$ & $\surd$ \\
G049.265+00.311 & III & 49.2649 & 0.3108 & s-pbhmi & 3.49(0.0) & 56.019 & 10.67 & $\surd$ & $\surd$ & $\surd$ \\
G049.544-00.883 & III & 49.5435 & -0.8831 & s-u-hmi & 2.71(0.0) & 28.815 & 7.73 & $\surd$ & {} & {} \\
G050.779+00.152 & III & 50.7786 & 0.1518 & spu-hmi & 3.29(4.84) & 8.88 & 7.42 & $\surd$ & $\surd$ & $\surd$ \\
G052.844-00.868 & III & 52.8438 & -0.8675 & s-u-smi & 3.94(4.33) & 9.306 & 6.25 & $\surd$ & {} & {} \\
G288.962+00.265 & III & 288.962 & 0.2645 & s-u-smi & 3.72(3.86) & 12.297 & 6.23 & $\surd$ & $\surd$ & {} \\
G289.944-00.891 & III & 289.9441 & -0.8907 & spu-smi & 11.6(2.01) & 125.42 & 8.57 & $\surd$ & $\surd$ & {} \\
G304.665-00.965 & III & 304.6652 & -0.9651 & s-ubhmi & 3.56(0.11) & 224.894 & 9.77 & $\surd$ & $\surd$ & {} \\
G305.475-00.096 & III & 305.4749 & -0.0962 & s-pbsmi & 2.89(0.0) & 11.977 & 6.54 & $\surd$ & $\surd$ & $\surd$ \\
G312.097-00.236 & III & 312.0971 & -0.2359 & s-u-hmi & 3.33(1.5) & 52.489 & 8.09 & $\surd$ & {} & {} \\
G318.471-00.214 & III & 318.4714 & -0.2141 & s-ubsmi & 2.53(0.84) & 12.298 & 10.01 & $\surd$ & $\surd$ & $\surd$ \\
G319.323-00.135 & III & 319.323 & -0.1349 & s-ubhmi & 2.99(0.0) & 17.285 & 11.15 & $\surd$ & $\surd$ & {} \\
G320.360-00.282 & III & 320.3597 & -0.2816 & s-pbhmi & 1.6(0.0) & 7.6 & 8.7 & $\surd$ & $\surd$ & {} \\
G320.517-00.340 & III & 320.5169 & -0.3397 & s-ubhmi & 3.03(2.44) & 6.47 & 8.71 & $\surd$ & {} & {} \\
G324.923-00.568 & III & 324.9233 & -0.5682 & s-ubsmi & 2.99(0.0) & 14.821 & 4.4 & $\surd$ & $\surd$ & $\surd$ \\
G327.119+00.511 & III & 327.1195 & 0.5108 & s-u-smi & 3.11(0.0) & 279.123 & 4.96 & $\surd$ & $\surd$ & $\surd$ \\
G327.402+00.445 & III & 327.402 & 0.4449 & s-pbsmi & 2.78(0.0) & 5.324 & 4.63 & $\surd$ & $\surd$ & $\surd$ \\
G327.654+00.126 & III & 327.6537 & 0.1255 & s-u-hmi & 2.36(0.0) & 11.033 & 5.16 & $\surd$ & $\surd$ & {} \\
G328.164+00.587 & III & 328.1644 & 0.5868 & s-u-smi & 3.24(0.0) & 31.103 & 8.98 & $\surd$ & $\surd$ & $\surd$ \\
G329.066-00.308 & III & 329.066 & -0.3079 & s-pbhmi & 6.23(7.12) & 106.091 & 11.62 & $\surd$ & $\surd$ & $\surd$ \\
G331.710+00.603 & III & 331.7095 & 0.603 & s-u-smi & 1.94(0.0) & 21.389 & 4.05 & $\surd$ & $\surd$ & $\surd$ \\
G333.314+00.105 & III & 333.3145 & 0.1051 & s-pbhmi & 4.86(0.0) & 60.606 & 11.94 & $\surd$ & $\surd$ & $\surd$ \\
G333.315+00.105 & III & 333.3146 & 0.1054 & s-pbhmi & 4.86(0.0) & 60.606 & 11.94 & $\surd$ & $\surd$ & $\surd$ \\
G337.052-00.226 & III & 337.052 & -0.2259 & s-u-smi & 2.84(0.56) & 71.454 & 10.86 & $\surd$ & {} & $\surd$ \\
G338.000-00.149 & III & 337.9999 & -0.1494 & s-u-smi & 3.4(0.46) & 395.611 & 11.33 & $\surd$ & {} & {} \\
G339.621-00.121 & III & 339.6212 & -0.1206 & s-u-hmi & 2.82(0.0) & 44.576 & 2.6 & $\surd$ & $\surd$ & $\surd$ \\
G339.986-00.425 & III & 339.9863 & -0.4247 & s-u-hmi & 4.22(0.0) & 11.245 & 10.53 & $\surd$ & {} & $\surd$ \\
G340.248-00.372 & III & 340.2483 & -0.3717 & s-u-hmi & 4.69(0.0) & 31.053 & 11.99 & $\surd$ & $\surd$ & $\surd$ \\
G342.958-00.318 & III & 342.9584 & -0.3177 & sp--smi & 4.61(4.21) & 71.55 & 12.7 & $\surd$ & $\surd$ & {} \\
G344.915-00.229 & III & 344.9149 & -0.2291 & s-ubsmi & 7.52(5.63) & 5.367 & 10.72 & $\surd$ & $\surd$ & {} \\
G349.883+00.231 & III & 349.8834 & 0.2306 & s-u-hmi & 4.52(1.89) & 417.632 & 21.77 & $\surd$ & {} & $\surd$ \\
G356.662-00.264 & III & 356.662 & -0.2641 & spu-hmi & 3.41(2.04) & 123.559 & 6.75 & $\surd$ & $\surd$ & $\surd$ \\
G357.169-00.916 & III & 357.1692 & -0.9159 & sp--smi & 3.13(0.87) & 11.291 & 19.56 & $\surd$ & {} & {} \\
\enddata
\tablecomments{Catalog of WGOs as robust MYSOs. Column (1): source name (Galactic coordinate); Column (2): group; Column (3) and (4): Galactic coordinate (longitude: l, latitude: b) of the 4.6 $\mu$m emission; Column (5): name of beat-fit model set; Column (6): protostellar mass estimated according to the method provided by \cite{1996MNRAS.281..257T}, uncertainty in brackets is obtained by error propagation; Column (7): accretion rate obtained from SED fitting \citep{Robitaille2017}; Column (8): distance obtained from \cite{2021MNRAS.504.2742E}; Column (9)-(10): thresholds indicating whether a source can form massive stars, and high-mass sources are marked with ``$\surd$"; Column (11): Class II methanol maser, and if the source is associated with it, then marked with ``$\surd$".}
\end{deluxetable*}
\renewcommand\arraystretch{7}
\setlength{\tabcolsep}{4pt}
\setlength{\LTleft}{0pt}

\startlongtable
\begin{deluxetable*}{ccccccccccc}
\tablenum{3}
\tablecaption{Candidate MYSOs \label{tab3}}
\tablewidth{0pt}
\tabletypesize{\scriptsize}
\tablehead{
\colhead{Name}  & \colhead{Group} & 
\colhead{$l$} & \colhead{$b$} & \colhead{Model set} & \colhead{Stellar mass} &
\colhead{Accretion rate} & \colhead{Distance} & \colhead{$\geqslant 870\ M_{\odot} R^{1.33}$} & \colhead{$\geqslant 1~\rm g/cm^{2}$} & \colhead{Class II}\\
{}& {} & {} & 
{} &{}& {($M_{\star}$)} &{($\dot M_{\rm acc}$)} & {}  & {} & {} & {methanol maser}\\
\colhead{}& {} & \dcolhead{^{\circ}} & 
\dcolhead{^{\circ}} &\colhead{}& \dcolhead{M_{\odot}} &\dcolhead{\times 10^{-4} M_{\odot}\rm yr^{-1}} & \colhead{kpc}  & \colhead{} & \colhead{} & \colhead{}} 
\decimalcolnumbers
\startdata
G001.934-00.170 & I & 1.9344 & -0.1699 & s-pbhmi & 2.61(0.12) & 3.256 & 10.09 & $\surd$ & {} & {} \\
G008.683-00.368 & I & 8.6833 & -0.3678 & s-u-hmi & 3.11(0.0) & 1.89 & 4.19 & $\surd$ & $\surd$ & $\surd$ \\
G010.477-00.358 & I & 10.4768 & -0.3584 & s-u-smi & 1.39(0.53) & 5.681 & 17.1 & $\surd$ & $\surd$ & {} \\
G011.918-00.613 & I & 11.9181 & -0.6131 & s---smi & 3.06(1.44) & 1.243 & 3.4 & $\surd$ & $\surd$ & {} \\
G012.683-00.183 & I & 12.6827 & -0.1829 & s-pbsmi & 2.72(1.73) & 1.507 & 2.4 & $\surd$ & $\surd$ & $\surd$ \\
G012.890+00.491 & I & 12.8904 & 0.4915 & s-pbhmi & 1.76(0.31) & 1.904 & 2.3 & $\surd$ & $\surd$ & $\surd$ \\
G028.832-00.252 & I & 28.832 & -0.2522 & s-u-smi & 3.17(0.32) & 4.383 & 4.86 & $\surd$ & $\surd$ & $\surd$ \\
G035.133-00.745 & I & 35.1326 & -0.745 & s-u-hmi & 2.05(0.0) & 2.578 & 2.08 & $\surd$ & $\surd$ & $\surd$ \\
G037.042-00.033 & I & 37.0418 & -0.0334 & s-u-hmi & 2.61(0.04) & 3.731 & 4.93 & $\surd$ & $\surd$ & {} \\
G058.774+00.645 & I & 58.7738 & 0.6449 & s-u-smi & 2.49(1.71) & 1.156 & 3.3 & $\surd$ & $\surd$ & $\surd$ \\
G083.465+00.157 & I & 83.4647 & 0.1574 & s-ubhmi & 1.97(1.27) & 2.106 & 2.88 & $\surd$ & {} & {} \\
G111.877+00.996 & I & 111.8773 & 0.9956 & s-ubhmi & 2.25(2.46) & 1.144 & 3.2 & $\surd$ & $\surd$ & {} \\
G294.512-01.620 & I & 294.5117 & -1.6205 & spu-hmi & 2.33(1.89) & 3.063 & 2.2 & $\surd$ & $\surd$ & $\surd$ \\
G305.536+00.942 & I & 305.5358 & 0.942 & s-u-hmi & 1.07(0.33) & 1.055 & 7.66 & $\surd$ & $\surd$ & {} \\
G309.535-00.739 & I & 309.5347 & -0.7389 & s-ubsmi & 4.06(3.58) & 1.756 & 4.58 & $\surd$ & $\surd$ & {} \\
G312.108+00.262 & I & 312.1084 & 0.2623 & s-ubhmi & 1.81(0.26) & 2.596 & 3.69 & $\surd$ & $\surd$ & $\surd$ \\
G320.184+00.840 & I & 320.184 & 0.8399 & s-pbhmi & 1.64(0.38) & 1.247 & 2.48 & {} & {} & {} \\
G321.936-00.006 & I & 321.9356 & -0.0063 & s-u-hmi & 1.71(0.0) & 2.485 & 2.13 & $\surd$ & $\surd$ & {} \\
G326.858-00.677 & I & 326.8583 & -0.677 & s-u-hmi & 1.45(0.0) & 2.579 & 3.87 & $\surd$ & $\surd$ & $\surd$ \\
G329.556+00.178 & I & 329.5555 & 0.1781 & s-pbsmi & 2.19(0.25) & 1.945 & 8.85 & $\surd$ & {} & $\surd$ \\
G331.623+00.526 & I & 331.623 & 0.5263 & s-u-hmi & 2.5(0.27) & 1.233 & 3.23 & $\surd$ & $\surd$ & {} \\
G332.122+00.936 & I & 332.1215 & 0.9358 & s-pbhmi & 1.61(0.0) & 1.47 & 4.36 & $\surd$ & $\surd$ & {} \\
G333.076-00.559 & I & 333.076 & -0.559 & s-pbsmi & 2.31(0.57) & 1.743 & 3.8 & $\surd$ & $\surd$ & {} \\
G333.465-00.163 & I & 333.4653 & -0.163 & s-u-hmi & 4.9(7.205) & 2.016 & 2.86 & $\surd$ & $\surd$ & $\surd$ \\
G334.332+00.964 & I & 334.3317 & 0.964 & s-ubsmi & 2.61(1.58) & 1.292 & 3.37 & $\surd$ & $\surd$ & {} \\
G335.789+00.174 & I & 335.7889 & 0.1743 & s-ubsmi & 3.44(0.15) & 3.191 & 3.23 & $\surd$ & $\surd$ & $\surd$ \\
G340.970-01.022 & I & 340.9701 & -1.0217 & s-pbsmi & 2.22(0.28) & 2.174 & 2.0 & $\surd$ & $\surd$ & $\surd$ \\
G341.238-00.271 & I & 341.2378 & -0.2708 & s-u-smi & 2.71(0.63) & 2.06 & 3.47 & $\surd$ & $\surd$ & $\surd$ \\
G343.501-00.473 & I & 343.5013 & -0.4731 & s-u-smi & 1.67(0.27) & 2.697 & 2.0 & $\surd$ & $\surd$ & $\surd$ \\
G349.644-01.095 & I & 349.6436 & -1.0953 & s-u-hmi & 2.26(2.13) & 2.239 & 2.2 & $\surd$ & $\surd$ & {} \\
G081.662+00.551 & I & 81.6615 & 0.5506 & s-pbsmi & 1.99(0.995) & 2.377 & 2.92 & $\surd$ & {} & {} \\
G305.193-00.005 & I & 305.1931 & -0.0055 & s-pbhmi & 1.62(1.11) & 3.464 & 2.85 & $\surd$ & $\surd$ & {} \\
G003.347+00.441 & II & 3.3465 & 0.4414 & s-u-smi & 3.38(1.32) & 2.741 & 11.16 & $\surd$ & {} & {} \\
G005.910-00.544 & II & 5.9095 & -0.5442 & s-pbhmi & 1.54(0.18) & 1.115 & 3.27 & $\surd$ & $\surd$ & {} \\
G006.922-00.251 & II & 6.9219 & -0.2513 & s-ubsmi & 2.15(0.89) & 1.387 & 3.82 & $\surd$ & $\surd$ & {} \\
G007.993-00.269 & II & 7.9931 & -0.2691 & s-u-hmi & 3.11(0.0) & 3.18 & 11.79 & {} & {} & {} \\
G012.903-00.032 & II & 12.9032 & -0.0315 & s-u-hmi & 2.88(0.57) & 4.059 & 4.7 & $\surd$ & $\surd$ & $\surd$ \\
G013.097-00.145 & II & 13.0968 & -0.1447 & s-ubhmi & 1.91(0.58) & 3.475 & 4.05 & $\surd$ & $\surd$ & {} \\
G013.280+00.092 & II & 13.2801 & 0.0917 & s-ubsmi & 1.55(0.18) & 4.567 & 13.09 & $\surd$ & {} & {} \\
G014.012-00.175 & II & 14.0118 & -0.1754 & s-u-hmi & 1.45(0.0) & 2.346 & 3.72 & $\surd$ & $\surd$ & {} \\
G014.320-00.133 & II & 14.3205 & -0.1328 & s-u-hmi & 3.06(2.06) & 1.021 & 12.57 & $\surd$ & {} & {} \\
G018.234+00.654 & II & 18.2342 & 0.6536 & sp--hmi & 2.97(0.27) & 3.948 & 12.79 & $\surd$ & $\surd$ & {} \\
G018.246-00.472 & II & 18.2464 & -0.4719 & s-u-smi & 1.25(0.29) & 3.202 & 12.15 & $\surd$ & {} & {} \\
G018.256-00.253 & II & 18.2565 & -0.2527 & s-u-hmi & 2.86(2.84) & 1.058 & 4.46 & $\surd$ & $\surd$ & {} \\
G018.930+00.083 & II & 18.9305 & 0.0829 & s-pbhmi & 2.39(1.56) & 2.592 & 12.36 & $\surd$ & {} & {} \\
G019.365-00.029 & II & 19.3651 & -0.0293 & s-pbhmi & 2.12(0.42) & 3.027 & 2.22 & $\surd$ & $\surd$ & $\surd$ \\
G020.234+00.084 & II & 20.234 & 0.0844 & s-u-hmi & 1.89(0.33) & 5.141 & 11.12 & $\surd$ & $\surd$ & {} \\
G022.038+00.222 & II & 22.0381 & 0.2222 & s-pbhmi & 2.4(0.0) & 1.775 & 3.43 & $\surd$ & $\surd$ & $\surd$ \\
G023.227+00.163 & II & 23.2272 & 0.1629 & spu-hmi & 1.08(0.17) & 1.043 & 4.59 & $\surd$ & {} & {} \\
G023.569+00.011 & II & 23.5685 & 0.0109 & s-pbhmi & 3.07(2.15) & 1.298 & 5.66 & $\surd$ & $\surd$ & {} \\
G024.633+00.153 & II & 24.6334 & 0.1528 & s-pbsmi & 1.36(0.0) & 3.268 & 3.45 & $\surd$ & $\surd$ & $\surd$ \\
G024.639-00.030 & II & 24.6389 & -0.0304 & s-ubsmi & 2.13(0.48) & 5.635 & 15.99 & $\surd$ & {} & {} \\
G024.942+00.074 & II & 24.9423 & 0.074 & s-ubhmi & 1.64(0.0) & 2.142 & 2.81 & $\surd$ & $\surd$ & {} \\
G025.569-00.095 & II & 25.5689 & -0.0953 & s-u-smi & 1.23(0.0) & 3.875 & 9.85 & $\surd$ & {} & {} \\
G025.920-00.141 & II & 25.92 & -0.1408 & s-ubhmi & 2.59(1.46) & 4.678 & 9.22 & $\surd$ & {} & $\surd$ \\
G026.663+00.275 & II & 26.6631 & 0.2747 & s-u-smi & 2.65(0.82) & 2.279 & 8.87 & $\surd$ & {} & {} \\
G026.770-00.102 & II & 26.7703 & -0.1025 & spu-hmi & 1.58(0.31) & 4.535 & 9.53 & $\surd$ & {} & {} \\
G027.294-00.156 & II & 27.2939 & -0.156 & s-u-smi & 1.96(1.98) & 1.317 & 9.8 & $\surd$ & {} & {} \\
G027.956+00.114 & II & 27.956 & 0.1142 & spu-hmi & 1.45(0.37) & 5.466 & 13.15 & $\surd$ & {} & {} \\
G028.320-00.011 & II & 28.3197 & -0.0108 & spu-smi & 4.79(6.59) & 1.229 & 5.44 & $\surd$ & $\surd$ & $\surd$ \\
G029.277-00.128 & II & 29.2773 & -0.1283 & s-u-hmi & 2.13(0.0) & 3.703 & 3.63 & $\surd$ & $\surd$ & $\surd$ \\
G029.278-00.128 & II & 29.2782 & -0.1284 & s-u-hmi & 2.13(0.0) & 3.703 & 3.63 & $\surd$ & $\surd$ & $\surd$ \\
G029.325-00.515 & II & 29.3246 & -0.5146 & sp--hmi & 2.05(1.0) & 1.29 & 8.94 & {} & {} & {} \\
G029.363-00.551 & II & 29.3627 & -0.5512 & s---smi & 2.16(0.3) & 2.76 & 10.74 & $\surd$ & {} & {} \\
G029.780-00.260 & II & 29.7796 & -0.2597 & s-ubhmi & 1.17(0.27) & 2.875 & 8.93 & $\surd$ & $\surd$ & {} \\
G030.300-00.203 & II & 30.3001 & -0.2032 & s-u-smi & 2.67(2.04) & 1.011 & 8.71 & $\surd$ & {} & $\surd$ \\
G030.945+00.157 & II & 30.9447 & 0.1573 & s-u-hmi & 2.17(1.38) & 1.562 & 8.57 & $\surd$ & {} & $\surd$ \\
G034.575-00.000 & II & 34.5746 & 0.0 & s-u-smi & 1.15(0.0) & 4.275 & 9.3 & $\surd$ & $\surd$ & {} \\
G034.712-00.644 & II & 34.7121 & -0.6445 & sp--smi & 2.19(1.31) & 1.97 & 10.86 & $\surd$ & {} & {} \\
G034.994-00.045 & II & 34.9938 & -0.0446 & s-pbsmi & 1.79(1.44) & 1.109 & 10.84 & $\surd$ & {} & {} \\
G035.499-00.021 & II & 35.4989 & -0.0208 & sp--smi & 3.6(3.75) & 2.701 & 10.1 & $\surd$ & $\surd$ & {} \\
G037.359-00.074 & II & 37.3587 & -0.0739 & sp--hmi & 3.22(3.04) & 1.253 & 10.7 & $\surd$ & {} & {} \\
G037.753+00.197 & II & 37.753 & 0.1972 & s-ubhmi & 2.13(1.0) & 1.649 & 10.44 & $\surd$ & $\surd$ & {} \\
G037.763-00.215 & II & 37.7629 & -0.2151 & s-ubsmi & 2.9(2.38) & 4.474 & 9.58 & $\surd$ & $\surd$ & {} \\
G043.326-00.202 & II & 43.3258 & -0.2025 & s-pbsmi & 2.36(0.34) & 1.217 & 8.16 & $\surd$ & $\surd$ & {} \\
G044.097+00.160 & II & 44.0965 & 0.16 & s-ubhmi & 1.54(0.81) & 1.504 & 8.31 & $\surd$ & $\surd$ & {} \\
G046.246-00.773 & II & 46.2464 & -0.7735 & s-u-hmi & 1.27(0.0) & 6.582 & 7.95 & $\surd$ & {} & {} \\
G046.871-00.301 & II & 46.8713 & -0.3012 & s-u-hmi & 1.95(0.15) & 1.272 & 8.63 & $\surd$ & {} & {} \\
G049.381-00.184 & II & 49.3813 & -0.184 & spu-hmi & 2.73(1.63) & 2.579 & 7.25 & $\surd$ & {} & {} \\
G050.916-00.722 & II & 50.9159 & -0.7217 & s-ubhmi & 3.08(2.64) & 1.082 & 7.7 & $\surd$ & {} & {} \\
G052.341+00.324 & II & 52.3408 & 0.3242 & s-ubhmi & 1.38(0.42) & 3.745 & 6.26 & $\surd$ & {} & {} \\
G054.371-00.613 & II & 54.3709 & -0.6132 & s-pbsmi & 2.29(0.0) & 1.399 & 7.3 & $\surd$ & $\surd$ & $\surd$ \\
G060.621-00.700 & II & 60.621 & -0.6997 & s-u-hmi & 1.29(0.21) & 3.612 & 5.05 & $\surd$ & {} & {} \\
G063.115+00.341 & II & 63.1148 & 0.3414 & s-ubsmi & 2.38(0.33) & 1.433 & 2.2 & $\surd$ & $\surd$ & {} \\
G074.569+00.846 & II & 74.5692 & 0.8455 & s-u-hmi & 1.63(0.895) & 3.251 & 4.57 & $\surd$ & $\surd$ & {} \\
G080.315+01.332 & II & 80.3154 & 1.3317 & s-pbsmi & 1.83(0.59) & 3.594 & 6.32 & $\surd$ & {} & {} \\
G284.467-00.369 & II & 284.4673 & -0.3693 & s-u-hmi & 1.75(0.42) & 2.054 & 5.5 & $\surd$ & {} & {} \\
G289.035-00.037 & II & 289.0348 & -0.0368 & spu-smi & 1.24(0.58) & 1.365 & 7.44 & $\surd$ & {} & {} \\
G291.180-00.290 & II & 291.1799 & -0.2899 & sp--hmi & 2.87(0.0) & 2.127 & 7.34 & $\surd$ & $\surd$ & {} \\
G295.106-01.678 & II & 295.1065 & -1.6778 & sp--hmi & 2.16(0.11) & 1.576 & 2.2 & $\surd$ & $\surd$ & {} \\
G295.867-00.130 & II & 295.8669 & -0.1301 & spu-smi & 2.46(1.47) & 1.264 & 8.58 & $\surd$ & {} & {} \\
G301.473-00.225 & II & 301.4734 & -0.2254 & spu-hmi & 1.46(0.3) & 3.909 & 6.39 & $\surd$ & $\surd$ & {} \\
G303.566-00.854 & II & 303.5657 & -0.8536 & s-pbsmi & 3.15(3.87) & 1.135 & 11.27 & $\surd$ & {} & {} \\
G303.846-00.363 & II & 303.8458 & -0.3631 & s-ubhmi & 1.78(0.41) & 2.325 & 11.49 & $\surd$ & {} & $\surd$ \\
G304.022+00.292 & II & 304.0224 & 0.2924 & s-ubsmi & 3.4(3.16) & 4.44 & 4.67 & $\surd$ & $\surd$ & {} \\
G305.254+00.081 & II & 305.2539 & 0.0806 & s-pbhmi & 1.48(0.0) & 1.587 & 4.57 & $\surd$ & {} & {} \\
G305.823-00.114 & II & 305.8225 & -0.114 & s-ubhmi & 3.65(3.14) & 2.273 & 5.92 & $\surd$ & $\surd$ & $\surd$ \\
G309.173-00.010 & II & 309.1727 & -0.0099 & s-pbsmi & 1.44(0.26) & 1.972 & 9.53 & $\surd$ & $\surd$ & {} \\
G313.691+00.098 & II & 313.6911 & 0.098 & s---smi & 2.72(0.28) & 2.097 & 8.52 & $\surd$ & $\surd$ & {} \\
G313.758+00.254 & II & 313.7584 & 0.2539 & s-ubhmi & 2.42(1.83) & 1.473 & 8.02 & $\surd$ & {} & {} \\
G314.211+00.210 & II & 314.2112 & 0.21 & s-u-smi & 1.19(0.43) & 1.335 & 7.45 & $\surd$ & $\surd$ & {} \\
G317.873-01.054 & II & 317.8731 & -1.0539 & s-u-hmi & 2.03(0.83) & 2.097 & 3.47 & $\surd$ & {} & {} \\
G321.148-00.529 & II & 321.1483 & -0.5289 & s-ubsmi & 2.23(1.32) & 1.482 & 3.84 & $\surd$ & $\surd$ & $\surd$ \\
G323.519-00.470 & II & 323.5186 & -0.4699 & s-u-hmi & 1.34(0.0) & 4.662 & 11.39 & $\surd$ & {} & {} \\
G326.314-00.481 & II & 326.3136 & -0.4807 & sp--hmi & 3.17(1.86) & 1.963 & 10.3 & $\surd$ & $\surd$ & {} \\
G326.920-00.021 & II & 326.9197 & -0.0213 & s-ubsmi & 2.06(1.18) & 1.03 & 3.32 & $\surd$ & {} & {} \\
G327.894+00.149 & II & 327.8945 & 0.1485 & s-ubhmi & 3.79(1.86) & 4.08 & 8.91 & $\surd$ & {} & {} \\
G328.140-00.432 & II & 328.1401 & -0.4317 & s-ubhmi & 1.71(0.27) & 1.753 & 2.73 & $\surd$ & $\surd$ & $\surd$ \\
G328.335-00.528 & II & 328.3349 & -0.5283 & s-ubhmi & 2.17(0.26) & 1.09 & 2.54 & $\surd$ & $\surd$ & {} \\
G329.014+00.985 & II & 329.0144 & 0.9849 & s-u-hmi & 0.93(0.0) & 1.157 & 5.92 & $\surd$ & $\surd$ & {} \\
G329.422-00.162 & II & 329.422 & -0.1621 & s-pbsmi & 3.04(0.26) & 1.794 & 4.35 & $\surd$ & $\surd$ & {} \\
G329.705-00.343 & II & 329.7052 & -0.3429 & s-ubhmi & 2.66(1.24) & 1.959 & 11.44 & $\surd$ & $\surd$ & {} \\
G331.083-00.475 & II & 331.0832 & -0.4752 & s-pbhmi & 1.45(0.0) & 1.106 & 3.91 & $\surd$ & $\surd$ & {} \\
G331.434-00.283 & II & 331.4341 & -0.2833 & s-u-hmi & 1.46(0.35) & 4.843 & 4.87 & $\surd$ & $\surd$ & {} \\
G332.350-00.436 & II & 332.3502 & -0.4358 & spu-hmi & 1.22(0.61) & 1.073 & 3.1 & $\surd$ & $\surd$ & $\surd$ \\
G332.725-00.621 & II & 332.7251 & -0.6206 & s-u-hmi & 2.06(0.49) & 4.186 & 3.8 & $\surd$ & $\surd$ & $\surd$ \\
G332.813-00.700 & II & 332.8129 & -0.7004 & s-ubsmi & 2.69(2.31) & 5.18 & 3.8 & $\surd$ & $\surd$ & $\surd$ \\
G333.203-00.046 & II & 333.2026 & -0.0459 & s-u-smi & 1.18(0.39) & 1.016 & 3.0 & $\surd$ & $\surd$ & {} \\
G333.254-00.027 & II & 333.254 & -0.0267 & sp--hmi & 2.71(0.49) & 1.847 & 10.83 & $\surd$ & {} & {} \\
G335.426-00.240 & II & 335.426 & -0.2397 & s-ubhmi & 1.71(0.36) & 5.653 & 2.95 & $\surd$ & $\surd$ & $\surd$ \\
G337.258-00.101 & II & 337.2578 & -0.101 & s-u-hmi & 2.9(0.23) & 3.679 & 11.17 & $\surd$ & $\surd$ & $\surd$ \\
G337.632-00.078 & II & 337.6318 & -0.0785 & s-pbhmi & 2.48(2.13) & 4.189 & 3.61 & $\surd$ & $\surd$ & $\surd$ \\
G337.697+00.381 & II & 337.6973 & 0.3814 & s---smi & 2.23(0.25) & 2.056 & 10.76 & $\surd$ & {} & {} \\
G338.325-00.409 & II & 338.3251 & -0.409 & s-u-hmi & 1.92(1.1) & 2.334 & 2.91 & $\surd$ & $\surd$ & $\surd$ \\
G338.467-00.010 & II & 338.4669 & -0.0102 & s-u-hmi & 1.41(0.52) & 2.085 & 5.94 & $\surd$ & $\surd$ & {} \\
G340.768-01.013 & II & 340.7675 & -1.0128 & s-ubhmi & 1.68(0.0) & 1.623 & 2.0 & $\surd$ & $\surd$ & {} \\
G340.822-01.028 & II & 340.8221 & -1.0285 & s-pbhmi & 2.6(0.19) & 1.136 & 2.0 & $\surd$ & {} & {} \\
G342.061+00.421 & II & 342.0605 & 0.4205 & spu-hmi & 2.94(0.21) & 3.068 & 4.51 & $\surd$ & $\surd$ & {} \\
G342.821+00.383 & II & 342.8208 & 0.3825 & s-u-hmi & 1.28(0.23) & 4.607 & 10.82 & $\surd$ & $\surd$ & {} \\
G343.527-00.505 & II & 343.5272 & -0.5053 & s-pbhmi & 1.72(1.02) & 1.854 & 3.04 & $\surd$ & $\surd$ & {} \\
G343.904-00.671 & II & 343.9037 & -0.6709 & s-ubhmi & 2.35(1.1) & 2.086 & 2.0 & $\surd$ & $\surd$ & {} \\
G344.103-00.661 & II & 344.1031 & -0.6613 & s-u-smi & 1.46(0.0) & 1.548 & 2.0 & $\surd$ & $\surd$ & {} \\
G346.355+00.105 & II & 346.3553 & 0.105 & s-pbsmi & 1.34(0.0) & 1.73 & 5.91 & $\surd$ & $\surd$ & {} \\
G348.003+00.192 & II & 348.0034 & 0.1916 & sp--hmi & 1.51(0.17) & 1.836 & 5.89 & $\surd$ & {} & {} \\
G350.749+00.505 & II & 350.7495 & 0.505 & s-ubsmi & 2.6(1.82) & 3.002 & 4.29 & $\surd$ & $\surd$ & {} \\
G354.688+00.508 & II & 354.6877 & 0.5083 & s-pbhmi & 1.26(0.0) & 1.139 & 4.11 & $\surd$ & $\surd$ & {} \\
G355.738+00.390 & II & 355.7384 & 0.39 & s-ubsmi & 2.37(1.35) & 1.615 & 9.65 & $\surd$ & {} & {} \\
G356.285+00.208 & II & 356.2851 & 0.2083 & s-u-hmi & 1.94(1.01) & 2.805 & 9.59 & $\surd$ & {} & {} \\
G356.344-00.070 & II & 356.3436 & -0.0705 & s-pbsmi & 1.15(0.5) & 1.609 & 6.81 & $\surd$ & $\surd$ & {} \\
G357.523+00.195 & II & 357.5234 & 0.1946 & sp--smi & 2.71(1.56) & 1.814 & 6.71 & $\surd$ & $\surd$ & {} \\
G358.370-00.468 & II & 358.3703 & -0.4675 & s-u-smi & 1.42(0.07) & 1.343 & 1.63 & $\surd$ & $\surd$ & $\surd$ \\
G358.808-00.086 & II & 358.8083 & -0.0856 & s-pbhmi & 1.94(1.31) & 2.087 & 7.76 & $\surd$ & $\surd$ & $\surd$ \\
G011.904-00.141 & III & 11.9037 & -0.1413 & s-ubsmi & 2.04(0.0) & 3.105 & 3.67 & $\surd$ & $\surd$ & $\surd$ \\
G018.661+00.038 & III & 18.6608 & 0.0376 & s-u-smi & 3.52(2.66) & 2.928 & 10.88 & $\surd$ & $\surd$ & {} \\
G024.114-00.174 & III & 24.1138 & -0.1745 & sp--smi & 4.75(7.18) & 1.079 & 4.69 & $\surd$ & $\surd$ & {} \\
G025.517-00.206 & III & 25.5171 & -0.2061 & s-u-hmi & 3.8(3.7) & 2.919 & 8.78 & $\surd$ & {} & {} \\
G028.147-00.004 & III & 28.1467 & -0.0044 & s-ubsmi & 3.63(3.86) & 2.158 & 5.4 & $\surd$ & $\surd$ & $\surd$ \\
G030.446-00.359 & III & 30.4455 & -0.3591 & sp--smi & 2.2(0.57) & 4.669 & 8.54 & {} & {} & {} \\
G031.496+00.177 & III & 31.4959 & 0.1771 & spu-smi & 2.46(2.42) & 5.388 & 8.21 & $\surd$ & $\surd$ & {} \\
G040.622-00.138 & III & 40.6221 & -0.1381 & s-u-hmi & 3.75(4.72) & 1.705 & 2.06 & $\surd$ & $\surd$ & $\surd$ \\
G053.331+00.040 & III & 53.3307 & 0.0399 & s-pbhmi & 2.76(2.42) & 2.875 & 8.31 & {} & {} & {} \\
G059.436+00.820 & III & 59.4364 & 0.8196 & s-pbsmi & 3.23(4.71) & 3.753 & 6.28 & $\surd$ & {} & $\surd$ \\
G063.904-00.061 & III & 63.9038 & -0.0605 & s-ubsmi & 2.95(2.48) & 1.052 & 8.28 & $\surd$ & {} & {} \\
G078.377+01.020 & III & 78.3769 & 1.0198 & s-pbhmi & 2.22(0.81) & 2.615 & 3.56 & $\surd$ & $\surd$ & {} \\
G087.556+00.502 & III & 87.5562 & 0.5016 & s-pbsmi & 1.65(0.83) & 1.034 & 6.3 & $\surd$ & {} & {} \\
G303.991+00.209 & III & 303.9912 & 0.2088 & s-pbhmi & 2.02(0.0) & 1.021 & 4.66 & $\surd$ & {} & {} \\
G304.367-00.336 & III & 304.3673 & -0.3361 & s-ubsmi & 3.3(0.0) & 4.594 & 11.95 & $\surd$ & {} & $\surd$ \\
G310.077-00.228 & III & 310.0767 & -0.2277 & spu-smi & 1.8(0.5) & 2.879 & 4.25 & $\surd$ & {} & {} \\
G314.199+00.157 & III & 314.1992 & 0.1571 & s-ubsmi & 2.0(1.26) & 3.222 & 7.57 & $\surd$ & {} & {} \\
G322.385+00.533 & III & 322.3851 & 0.533 & s-u-hmi & 1.96(0.04) & 1.555 & 3.36 & $\surd$ & $\surd$ & {} \\
G322.521+00.637 & III & 322.5212 & 0.6366 & s-pbhmi & 2.02(0.0) & 3.043 & 3.37 & $\surd$ & $\surd$ & {} \\
G328.549+00.272 & III & 328.549 & 0.2716 & spu-smi & 3.93(6.16) & 3.676 & 3.53 & $\surd$ & $\surd$ & {} \\
G329.610+00.114 & III & 329.6098 & 0.1137 & s-u-smi & 6.13(10.03) & 1.59 & 3.8 & $\surd$ & $\surd$ & $\surd$ \\
G337.110+00.348 & III & 337.1098 & 0.3483 & spu-smi & 2.83(2.5) & 4.981 & 4.97 & $\surd$ & {} & {} \\
G338.737+00.175 & III & 338.7372 & 0.1747 & s-u-hmi & 2.44(1.92) & 1.078 & 3.29 & $\surd$ & {} & {} \\
G340.746-01.002 & III & 340.7458 & -1.0024 & s-pbhmi & 1.83(0.0) & 1.963 & 2.0 & $\surd$ & $\surd$ & {} \\
G340.937-00.231 & III & 340.9375 & -0.2308 & spu-hmi & 1.21(0.02) & 4.359 & 3.51 & $\surd$ & $\surd$ & {} \\
G341.990-00.102 & III & 341.9905 & -0.1019 & s-pbhmi & 1.84(0.58) & 1.5 & 3.54 & $\surd$ & {} & $\surd$ \\
G343.721-00.223 & III & 343.7213 & -0.2225 & s-pbsmi & 2.59(2.08) & 1.139 & 2.0 & $\surd$ & $\surd$ & {} \\
G352.158+00.402 & III & 352.1579 & 0.4025 & sp--hmi & 1.97(0.67) & 3.006 & 2.0 & $\surd$ & $\surd$ & {} \\
G356.596-00.503 & III & 356.5962 & -0.5028 & sp--smi & 3.82(3.94) & 4.177 & 12.86 & $\surd$ & $\surd$ & {} \\
G357.533+00.236 & III & 357.5325 & 0.2363 & s-pbhmi & 2.12(0.86) & 3.0 & 6.66 & $\surd$ & $\surd$ & {} \\
\enddata
\tablecomments{Catalog of WGOs as Candidate MYSOs.}
%\tablecomments{Catalog of WGOs as robust MYSOs. Column (1): source name (Galactic coordinate); Column (2): group; Column (3) and (4): Galactic coordinate (longitude: l, latitude: b) of the 4.6 $\mu$m emission; Column (5): name of beat-fit model set; Column (6): protostellar mass estimated according to the method provided by \cite{1996MNRAS.281..257T}, uncertainty in brackets is obtained by error propagation; Column (7): accretion rate obtained from SED fitting \citep{Robitaille2017}; Column (8): distance obtained from \cite{2021MNRAS.504.2742E}; Column (9)-(10): thresholds indicating whether a source can form massive stars, and high-mass sources are marked with ``$\surd$"; Column (11): Class II methanol maser, and if the source is associated with it, then marked with ``$\surd$".}
\end{deluxetable*}
\acknowledgments

We would like to thank the anonymous referees for valuable comments which improved the quality of the paper. We would to thank Dr. Zhiwei CHEN from Purple Mountain Observatory, Chinese Academy of Sciences for useful discussions on this paper. This work is supported by the Ministry of Science and Technology of China through grant 2010DFA02710, the Key Project of International Cooperation, and by the National Natural Science Foundation of China through grants 11503035, 11573036, 11373009, 11433008, 11403040 and 11403041. Guoyin ZHANG acknowledges support from China Postdoctoral Science Foundation (No.2021T140672), and National Natural Science foundation of China (No.U2031118).

\vspace{5mm}
%\facilities{WISE}

\bibliography{WGOs_cat}

\end{document}